\documentclass[10pt]{article} \usepackage{emulateapj}

\newcommand{\FeII}{Fe$\;${\small\rm II}\relax}

\newcommand{\TiII}{Ti$\;${\small\rm II}\relax}
\newcommand{\htwo}{H$_2$}
\newcommand{\HI}{H$\;${\small\rm I}\relax}
\newcommand{\NHI}{$N({\rm H \; \mbox{\small\rm I}})$}
\newcommand{\NDI}{$N({\rm D \; \mbox{\small\rm I}})$}
\newcommand{\NOI}{$N({\rm O \; \mbox{\small\rm I}})$}
\newcommand{\NNI}{$N({\rm N \; \mbox{\small\rm I}})$}
\newcommand{\DI}{D$\;${\small\rm I}\relax}
\newcommand{\OI}{O$\;${\small\rm I}\relax}

\newcommand{\NI}{N$\;${\small\rm I}\relax}
\newcommand{\lya}{Ly$\alpha$}
\newcommand{\lyb}{Ly$\beta$}

\newcommand{\err}[2]{\ensuremath{^{+ #1}_{- #2}}}

\newcommand{\kms}{km~s$^{-1}$\relax}

\newcommand{\fuse}{{\em FUSE}\relax}
\newcommand{\iue}{{\em IUE}\relax}
\newcommand{\imaps}{{\em IMAPS}\relax}
\newcommand{\hst}{{\em HST}\relax}
\newcommand{\copernicus}{{\em Copernicus}\relax}
\newcommand{\wmap}{{\em WMAP}\relax}

\received{2005 June 16}
\begin{document}

\title{The Deuterium, Oxygen, and Nitrogen Abundance Toward LSE 44}

\author{S.D.  Friedman\altaffilmark{1},
G. H\'ebrard\altaffilmark{2},
T.M.  Tripp\altaffilmark{3},
P. Chayer\altaffilmark{4},
K. R. Sembach\altaffilmark{1}}

\altaffiltext{1}{Space Telescope Science Institute, Baltimore, MD 21218;
friedman@stsci.edu}
	
\altaffiltext{2}{Institut d'Astrophysique de Paris, CNRS, 98 bis bld Arago,
F-75014 Paris, France}

\altaffiltext{3}{Department of Astronomy,
University of Massachusetts, Amherst, MA 01003}

\altaffiltext{4}{Department of Physics \& Astronomy, 
The Johns Hopkins University, Baltimore, MD 21218}

\begin{abstract} 

We present measurements of the column densities of interstellar \DI, \OI, \NI,
and \htwo\ made with the {\em Far Ultraviolet Spectroscopic Explorer} (\fuse),
and of \HI\ made with the {\em International Ultraviolet Explorer} (\iue) toward
the sdO star LSE 44 [{($l$,$b$) = (313\fdg37, +13\fdg49)}; d = $554\pm66$ pc; z
= $+129\pm15$ pc].  This target is among the seven most distant Galactic sight
lines for which these abundance ratios have been measured.  The \HI\ column
density was estimated by fitting the damping wings of interstellar \lya.  The
column densities of the remaining species were determined with profile fitting
analyses, and supplemented with curve of growth analyses for \OI\ and \htwo.  We
find log $N$(\DI) = $15.87\pm0.08$, log $N$(\OI) = $17.57\err{0.21}{0.15}$, log
$N$(\NI) = $16.43\pm0.14$, and log $N$(\HI) = $20.52\err{0.20}{0.36}$ (all
errors $2\,\sigma$).  This implies D/H = $(2.24\err{1.39}{1.32}) \times
10^{-5}$, D/O = $(1.99\err{1.30}{0.67}) \times 10^{-2}$, D/N =
$(2.75\err{1.19}{0.89}) \times 10^{-1}$, and O/H = $(1.13\err{0.96}{0.71})
\times 10^{-3}$.  Of the most distant Galactic sight lines for which the
deuterium abundance has been measured LSE 44 is one of the few with D/H higher
than the Local Bubble value, but D/O toward all these targets is below the Local
Bubble value and more uniform than the D/H distribution.

\end{abstract}


%%%%%%%%%%%%%%%%%%%%%%%%%%%%%%%%%%%%%%%%%%%%%%%%%%%%%%%%%%%%%%%%%%%%%%

\section{Introduction}

Precise measurements of primordial abundances of the light elements deuterium
(D), $^3$He, $^4$He, and $^7$Li relative to hydrogen have been a goal of
astronomers for many years.  In the standard Big Bang nucleosynthesis model,
these quantities are related in a straightforward way to the baryon-to-photon
ratio in the early universe, from which $\Omega_{\rm B}$, the fraction of the
critical density contributed by baryons, may be determined (Boesgard \& Steigman
1985).  Since deuterium is easily destroyed in stellar interiors (astration),
and no significant production mechanisms have been identified (Reeves et al.
1973; Epstein, Lattimer, \& Schramm 1976), the D/H ratio is expected to
monotonically decrease with time.

Precise measurements of D/H in the local interstellar medium (ISM) were first
made using \copernicus\ (Rogerson \& York 1973), \hst\ (Linsky et al.  1995),
and \imaps\ (Jenkins et al.  1999; Sonneborn et al.  2000).  In the last several
years many additional measurements of D/H, D/O, D/N, and related species, have
been made using \fuse\ (Moos et al.  2002; H\'ebrard \& Moos 2003; Wood et al.
2004).  These local ISM measurements represent a lower limit to the primordial
value.  When corrected for the effects of astration (Tosi et al.  1998), the
results should be comparable to those obtained in low-metallicity, high redshift
\lya\ clouds (Crighton et al.  2004; Kirkman et al.  2003), which should be very
nearly the primordial value itself.

As data on more sight lines has accumulated the situation is not as simple as
originally envisioned.  At high redshift, measurements of D/H vary from $(1.6-
4) \times 10^{-5}$ (Crighton et al.  2004; Pettini \& Bowen 2001), despite the
fact that the metallicity in all these environments is low enough to suggest
that D/H should match the primordial value.  More locally, it is now accepted
that for sight lines within the Local Bubble, a region of hot, low-density gas
(Sfeir et al.  1999) within a distance of $100-150$ pc and log \NHI $\la 19.2,$
in which the Sun lies, D/H is nearly constant, but the value is uncertain.
Based on measurements toward 16 targets, Wood et al.  (2004) state that
(D/H)$_{LB}=1.56\pm{0.04} \times 10^{-5}$.  Alternatively, (D/H)$_{LB}$ may be
inferred from measurements of D/O and O/H.  H\'ebrard \& Moos (2003) discuss the
advantages of this approach, including potential reduction of systematic errors
associated with measuring \NDI and \NHI\ directly, which differ by five orders
of magnitude.  Their approach relies on the observed high spatial uniformity of
O/H (Meyer, Jura, \& Cardelli 1998; Meyer 2001; Andr\'e et al.  2003).  The
result is (D/H)$_{LB}=1.32\pm{0.08} \times 10^{-5}$, about $2\sigma$ different
than the Wood et al.  value.

At distances greater than $\sim500$ pc, or more precisely, at log \NHI $\ga
20.5,$ the few sight lines measured have lower D/H.  Based on a suggestion
originally made by Jura (1982) and extended by Draine (2004a, 2004b) that under
the right thermodynamic conditions deuterium can be depleted onto dust grains,
Wood et al.  (2004) have interpreted D/H variations as being due to various
levels of depletion and, over the longer sight lines, the effects of averaging
the D/H concentration over multiple clouds along the sight line.  In this
scenario, the low value at large distances represents the gas phase abundance in
the local disk, (D/H)$_{LDg} = (0.85\pm0.09) \times 10^{-5},$ and one must
include the depleted deuterium to determine the total deuterium abundance.  An
alternate explanation is offered by H\'ebrard \& Moos (2003).  Based on D/O,
D/N, and D/H measurements, they suggest that the low gas-phase value observed at
high column densities and distances is the true Galactic disk value, and there
is not a significant reservoir of deuterium trapped in dust grains.  D/H has an
unusually high value in the Local Bubble if this interpretation is correct.  The
reason for this high value is not understood.  One possibility is the infall of
primordial, deuterium-rich material, and another is non-primordial deuterium
production, such as in stellar flares.  This is discussed further in \S5.

To test this hypothesis it is necessary to make additional measurements of \DI,
\HI, \OI, \NI, \htwo, and additional species that might yield information about
depletion and physical conditions, toward targets with log \NHI $\ga 20.5$ In
this paper we present the results of an analysis of the sight line toward LSE
44, an sdO star located at a distance of $554\pm66$ pc and well within this
column density regime.  We derive \DI, \OI, \NI, and \htwo\ column densities
with data obtained from the {\em Far Ultraviolet Spectroscopic Explorer} (\fuse)
and estimate the \HI\ column density with data from the {\em International
Ultraviolet Explorer} (\iue).

In \S2 we present a summary of the observations and a description of the data
reduction processes.  In \S3 the stellar properties of LSE 44 are discussed.
The details of the analysis of the \DI, \OI, \NI, and \htwo\ column densities
are given in \S4, and the \HI\ analysis is presented in \S 5.  In \S6 we discuss
the results of this study.

\section{Observations and Data Processing}

The \fuse\ instrument consists of four co-aligned spectrograph channels
designated LiF1, LiF2, SiC1, and SiC2, named for the coatings on their optics,
which were selected to optimize throughput.  Each channel illuminates a pair of
microchannel plate segments, labeled A and B.  The LiF channels have a bandpass
of approximately 1000-1187 \AA\, and the SiC channels approximately 910-1090
\,\AA.  Thus, each observed spectral line may appear in the spectra from
multiple channels.  The \fuse\ resolution is approximately 20 \kms\ (FWHM), but
it varies slightly between channels, and within a channel, as a function of
wavelength.  A detailed description of the \fuse\ instrumentation and
performance is given by Sahnow et al.  (2000).

LSE 44 was observed with \fuse\ for a total of 86 ksec under programs P2051602 -
P2051609.  A total of 119 exposures are in the MAST archive, obtained with
various levels of alignment success for the separate channels, and some were
obtained when one or both detectors were not at full voltage. The observation
log is shown in Table 1, and the total exposure time by channel is shown in
Table 2.  All data were obtained in time-tag mode with focal-plane offsets
(FP-SPLITS) in the MDRS ($20 \arcsec \times 4 \arcsec$) aperture.

The data were reduced using CALFUSE pipeline version 3.0.6.  We used the default
pixel size of 0.013\AA.  The signal-to-noise ratio per pixel in the continuum
region around 928\AA\ for both SiC1B and SiC2A spectra is typically about 3.5 in
individual, well-aligned exposures of $\sim900$s, the maximum duration of any
exposure.  The individual 1-dimensional spectra were co-added to form the final
spectrum for each channel and detector segment separately after removing the
relative shifts between individual spectra on the detector caused by image and
grating motion (Sahnow et al.  2000).  The shifts were determined by
cross-correlating the individual spectra over a limited wavelength range which
contained prominent spectral features but no airglow lines.  Typical shifts were
$\lesssim 4$ pixels, or $\lesssim 0.050$\AA.  The S/N ratios in the summed
spectra are approximately 26 and 27, respectively, for the SiC1B and SiC2A
channels.  Narrow interstellar absorption features in the summed spectra do not
exhibit unexpected broadening, indicating that the summing process did not smear
the data.  Although \fuse\ observations of \HI\ and \OI\ lines are sometimes
contaminated by airglow emission, data obtained during spacecraft orbital night
showed that this was unimportant in our analysis, so the combined day and night
data were used.

\begin{figure}
\epsscale{0.90}
\plotone{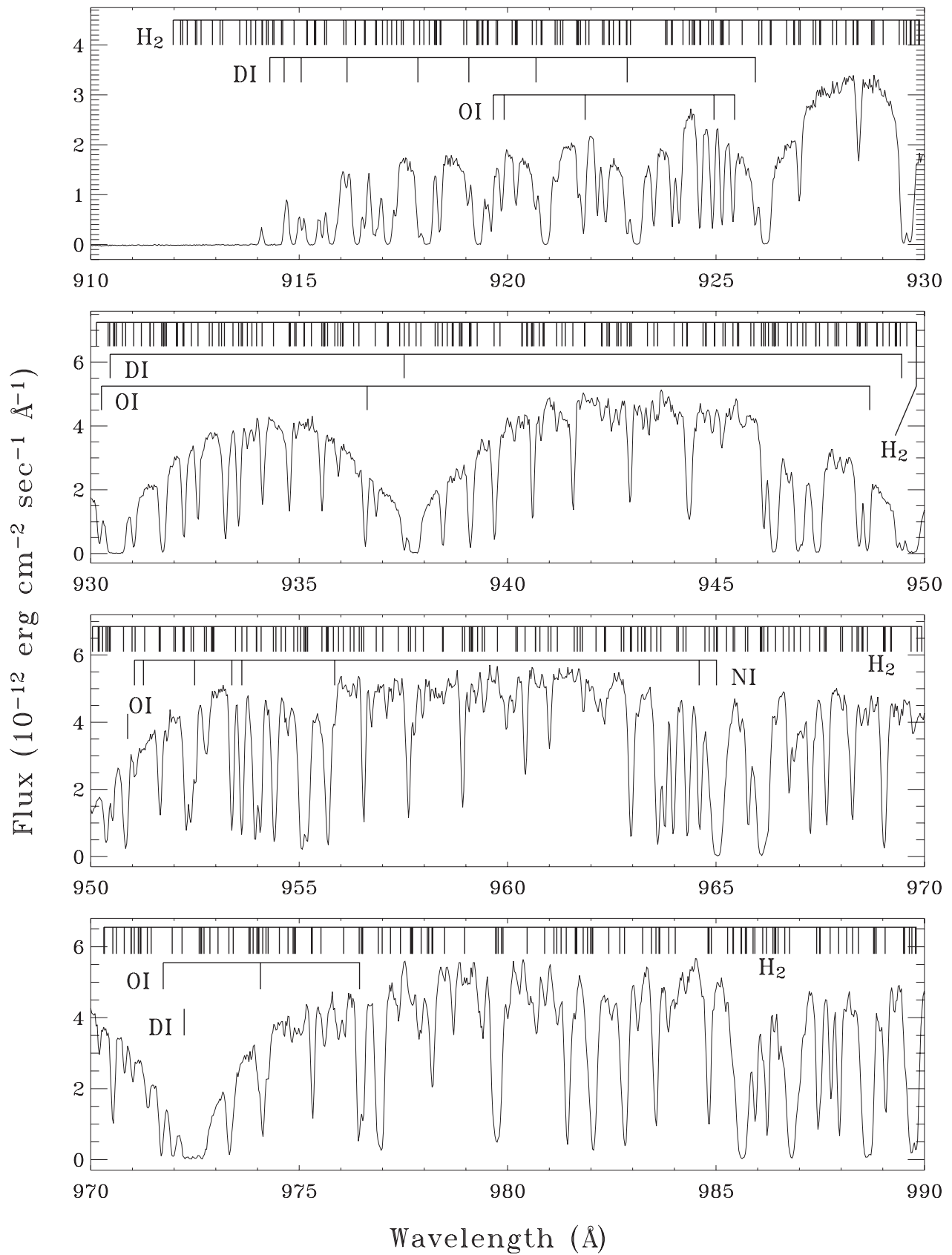}

\caption{The SiC1B spectrum of LSE 44.  The positions of interstellar \DI, \OI,
\NI, and \htwo\ absorption lines are indicated.  Many of the unmarked lines are
due to metals arising in the atmosphere of this subdwarf O star.  The data have
been binned by 2 pixels ($\sim 0.026$ \AA) for display purposes only in this
figure.  Note that the flux scale is not the same on all panels.}

\label{fig_s1bspec}
\end{figure}

Figure~\ref{fig_s1bspec} shows the co-added SiC1B spectrum of LSE~44, which
covers the wavelength range $915 \sim 990$\AA.  \DI, \OI, \NI, and \htwo\ lines
arising in the ISM are identified.

The \HI\ column density along this line of sight was estimated using the
Ly$\alpha$ profile from the single high resolution spectrum of LSE~44 taken with
\iue.  The 16.5 ksec observation was made on 29 August 1986 (see Table 1), and
reduced using IUESIPS.  The analysis of this data is discussed in more detail in
\S 4.4.

\section{Stellar Properties of LSE~44}

\begin{figure}
\plotone{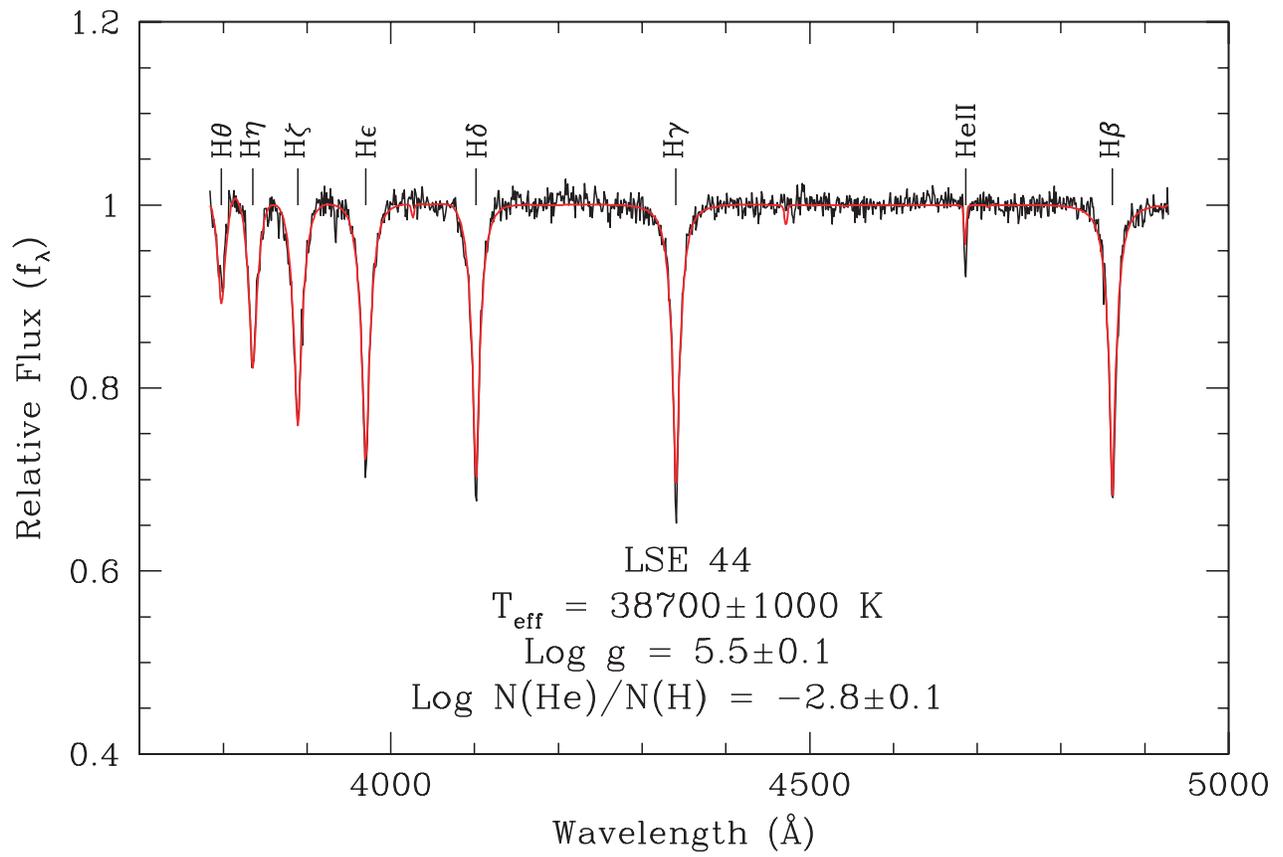}
\caption{{\it Black line:}  normalized optical spectrum of the sdO star
LSE~44.  {\it Red line:} Best model matching the optical spectrum.}
\label{fig:optical_spectrum}
\end{figure}

In his survey of subluminous O stars, Drilling (1983) obtained photometric and
spectroscopic data for LSE~44.  His $UBV$ photometry gives $V = 12.45$, $B-V =
-0.24$, and $U-B = -1.18$ (Table 3), and indicates that the star is relatively
hot.  He describes the optical spectrum of LSE~44 by saying that the only lines
present are the \ion{He}{2} $\lambda$4686 line and the Balmer series.  He adds
that the appearance of these lines is very similar to those in the spectrum of
the sdOB star Feige~110, but the Balmer lines are stronger and the \ion{He}{2}
$\lambda$4686 line is weaker in the spectrum of LSE~44 (see, e.g., Heber 1984;
Friedman et al.  2002).  In order to measure the atmospheric parameters of
LSE~44, D.  Kilkenny (2004, private communication) obtained an optical spectrum
of LSE~44 at the South African Astronomical Observatory with the 1.9-m telescope
and CCD spectrograph.  The optical spectrum covers the wavelength range 3400 --
5400 \AA\ with a resolution of about $\sim 3$~\AA.  The normalized spectrum is
plotted in Figure~\ref{fig:optical_spectrum}.  As reported by Drilling (1983),
our optical spectrum shows broad Balmer lines starting from H$\beta$ to
H$\theta$ and the \ion{He}{2} $\lambda$4686 line, which is the only helium line
detected in this spectrum.  The \ion{Ca}{2} K $\lambda$3935 line is also
detected in the optical spectrum.

The atmospheric parameters of LSE~44 were obtained by fitting the
optical spectrum illustrated in Figure~\ref{fig:optical_spectrum} with
a grid of synthetic spectra and by using the Marquardt method (see,
Bevington \& Robinson 1992).  This method optimizes the effective
temperature, gravity, and helium abundance in order to obtain a model
that best matches the observed spectrum.  We computed a grid of NLTE
H/He atmosphere models with 20,000~K~$\leq T_{\rm{eff}} \leq 80$,000~K
in steps of 2,000~K, for different values of the surface gravity $4.8
\leq \log g \leq 6.4$ in steps of 0.2 dex, and for different helium
abundances $-4.0 \leq \log N({\rm{He}})/N({\rm{H}}) \leq 0.0$ in steps
of 0.5 dex.  The models were computed with the stellar atmosphere
codes TLUSTY and SYNSPEC (see,
e.g., Hubeny \& Lanz 1995).  Figure~\ref{fig:optical_spectrum} shows
our best fit for the optical spectrum of LSE~44.  We obtain
$T_{\rm{eff}} = 38$,700$\pm1$,000~K, $\log g = 5.5\pm0.1$, and $\log
N({\rm{He}})/N({\rm{H}}) = -2.8\pm0.1$.  We used these atmospheric
parameters to compute the stellar Ly$\alpha$ line profile in order
to measure the \ion{H}{1} column density toward LSE~44. (see \S4.4).

\begin{figure}
\plotone{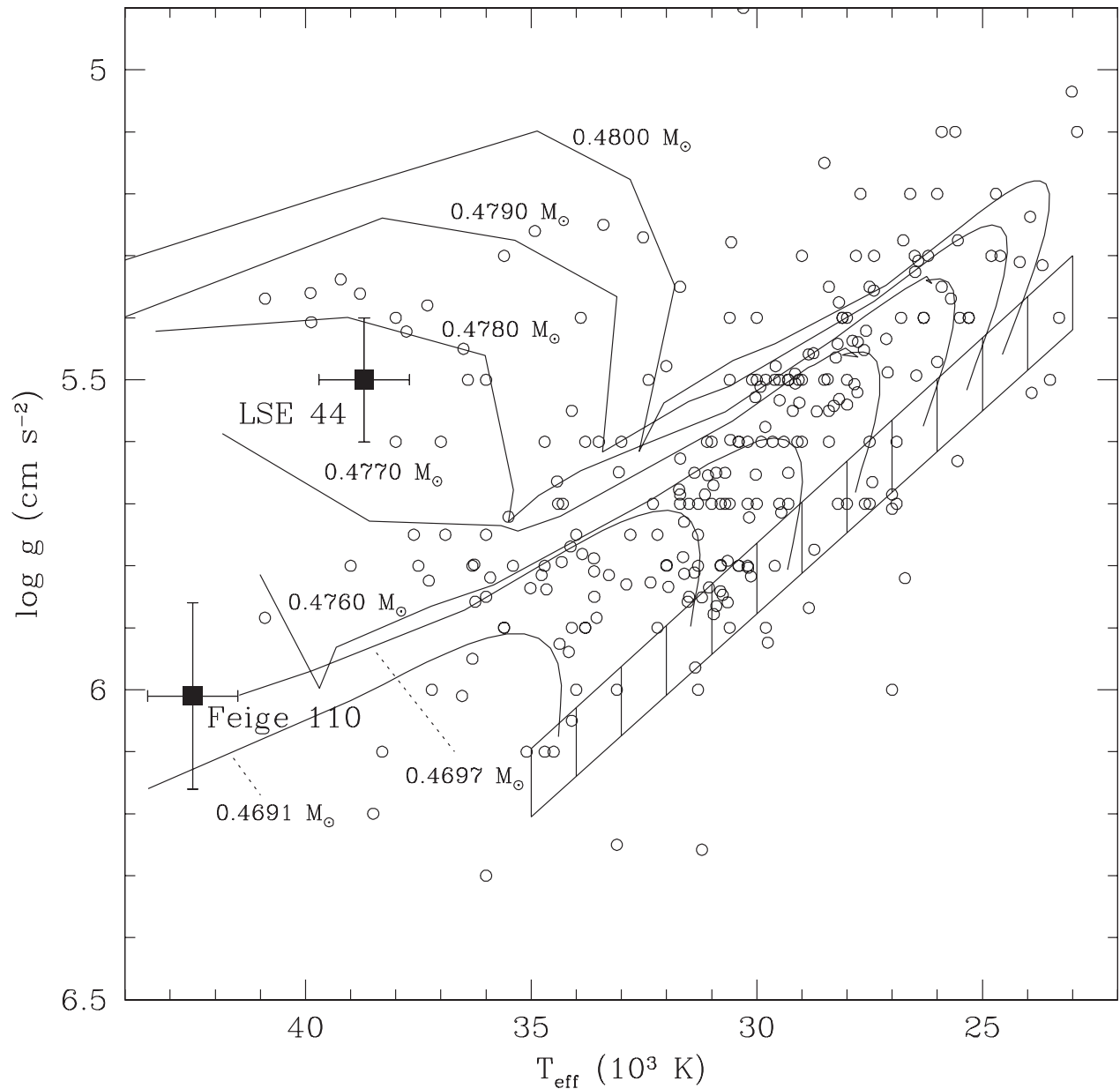}

\caption{Position of the sdO star LSE~44 on the $T_{\rm{eff}}$-$g$ plane.  Open
circles are parameters of observed sdB and sdOB stars.  The position of the sdOB
star Feige~110 is also indicated (Friedman et al.  2002).  Solid lines are
post-EHB evolutionary tracks from B.  Dorman (1999, private communication).  The
numbers indicate the mass of a star in solar masses that corresponds to an
evolutionary track.  The hashed region is the ZAEHB.}

\label{fig:teff_g_plane}
\end{figure}

The evolutionary status and physical properties of LSE~44 can be estimated by
comparing its position in a $T_{\rm{eff}}$-$g$ diagram with evolutionary
sequences.  Figure~\ref{fig:teff_g_plane} shows the position of LSE~44 in such a
diagram with the positions of a sample of hot subdwarf B stars (sdB).  It also
shows post-extreme horizontal branch (post-EHB) evolutionary sequences that were
computed by B.  Dorman (1999, private communication).  The position of LSE~44 in
the $T_{\rm{eff}}$-$g$ diagram suggests that it is an sdB star that has evolved
from the EHB.  Even though LSE~44 is classified as an sdO star based on its
optical spectrum, it relates to the sdB stars.  sdB stars have effective
temperatures and gravities in the ranges 24,000~K $\lesssim T_{\rm{eff}}
\lesssim 40$,000~K and $5.0 \lesssim \log g \lesssim 6.2$, and helium abundances
that are typically a factor of ten smaller than solar (see, e.g., Saffer et al.
1994).  The evolutionary sequences in Figure~\ref{fig:teff_g_plane} represent
the evolution of typical sdB stars that have helium-burning cores of $\sim
0.477$~M$_\odot$ and small hydrogen envelopes $M_{\rm{env}} \lesssim
0.0045$~M$_\odot$.  Each evolutionary track corresponds to models with about the
same helium core mass but with different hydrogen envelope masses.  In these
models the hydrogen envelopes are too small to ignite the hydrogen burning
shell, and therefore the stars appear subluminous, hot, and compact.  In
Figure~\ref{fig:teff_g_plane} the hatched region corresponds to the zero-age EHB
(ZAEHB).  In this way, an EHB star evolves from the ZAEHB and moves first to the
right and then to the left toward the white dwarf region without reaching the
asymptotic giant branch (AGB).  A star spends $\sim 10^8$ yr on the EHB and
$\sim 10^7$ yr off the EHB en route to the white dwarf region.

The spectral type sdO is a spectroscopic class that
includes stars showing strong Balmer lines and \ion{He}{2} lines.
Because of its high effective temperature and low helium abundance,
LSE~44 appears as an sdO star.  The evolutionary tracks surrounding the
position of LSE~44 show that stars in this portion of the
$T_{\rm{eff}}$-$g$ diagram have left the helium core burning phase and
undergo a helium shell burning phase.  One can assume that the mass of
LSE~44 is $\sim 0.4775$~M$_\odot$ and estimate its distance by using
the expression

\begin{equation}
D = \sqrt{\frac{GM}{g}\frac{4\pi\,H_{V}}{F_{0,V}}}\; 10^{0.2(V-A_V)},
\end{equation}

\noindent where $G$ is the gravitational constant, $M$ is the mass of
the star, $g$ is the gravitational acceleration, $H_\nu$ is the
Eddington flux weighted by the Johnson passband $V$ of Bessel (1990),
$F_{0,V}$ is the average absolute flux of Vega at $V$ as specified in
Heber et al.  (1984), $V$ is the apparent visual magnitude, and $A_V$
is the interstellar absorption.  By using the best model fit
parameters and the passbands of Bessel (1990) we computed the
intrinsic color index $(B-V)_0$ and obtained the color excess $E(B-V)
= 0.05\pm0.03$, which gives $A_V = 0.155\pm0.093$ when using $R_V =
3.1$.  The distance to LSE~44 is then $554\pm66$~pc.

\begin{figure}
\plotone{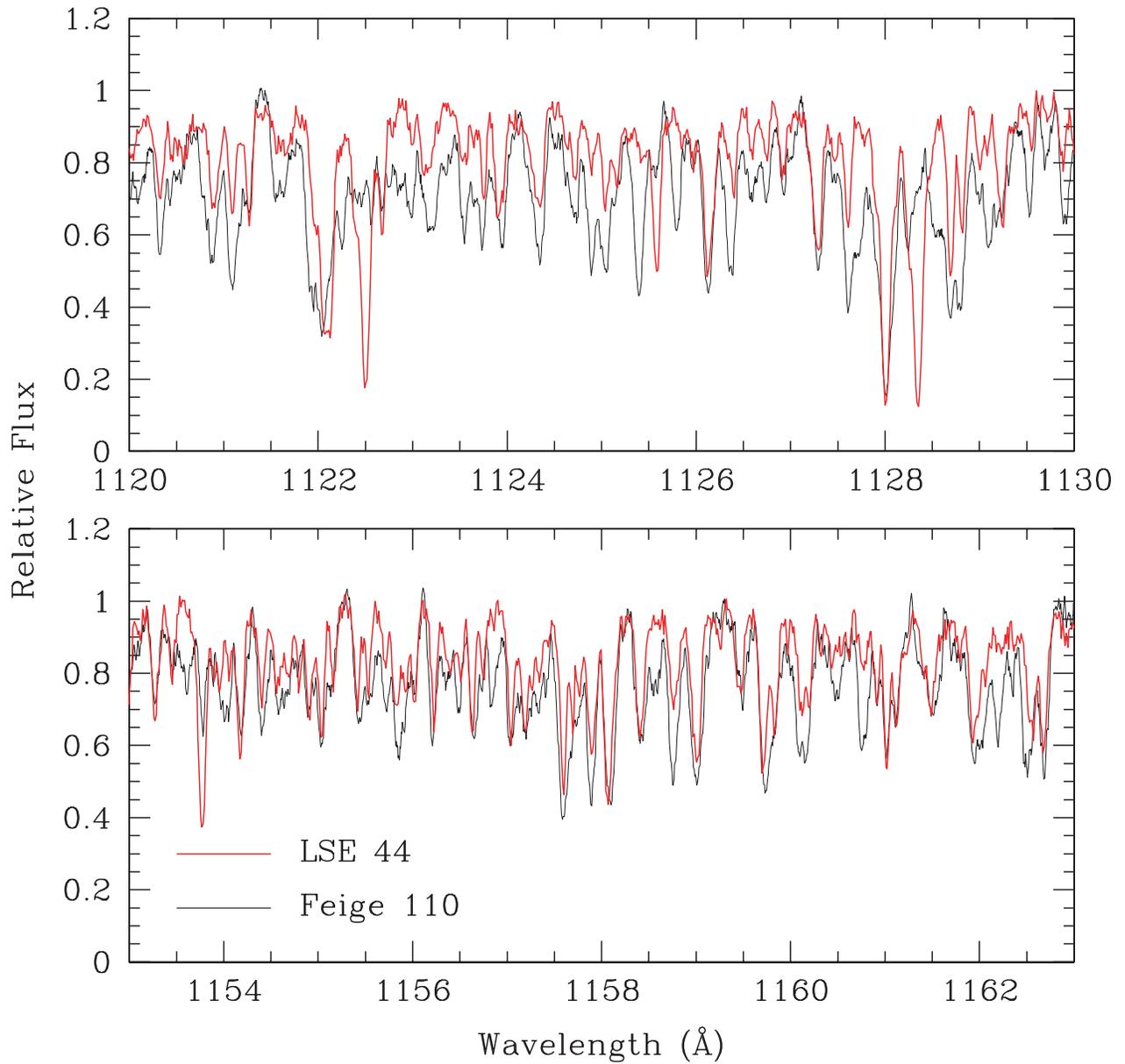}

\caption{Portions of the {\it FUSE} spectrum of LSE~44 ({\it red line}) compared
to the {\it FUSE} spectrum of the sdOB star Feige~110 ({\it black line}). Many
unidentified lines appear in spectra of both stars.}

\label{fig:lse44_feige110}
\end{figure}

The {\it FUSE} spectrum of LSE~44 shows many photospheric and interstellar
absorption lines.  The strongest photospheric lines are the Lyman series that
start at Ly$\beta$ and go up to lines close to the series limit.  Because of the
high gravity the photospheric Lyman lines merge together before reaching the
series limit, and the remaining Lyman lines close to this limit are mainly due
to interstellar \ion{H}{1}.  The other strong photospheric lines are \ion{He}{2}
$\lambda$1084, \ion{C}{3} $\lambda\lambda$1175, \ion{N}{3} $\lambda\lambda$1184,
\ion{N}{4} $\lambda\lambda$923 and $\lambda$955, \ion{Si}{4}
$\lambda\lambda$1126 and $\lambda$1066, \ion{P}{5} $\lambda\lambda$1123,
\ion{S}{4} $\lambda\lambda$1070, and \ion{S}{6} $\lambda\lambda$937.  There are
also many fainter unidentified stellar lines that depress the continuum.  For
instance, Figure~\ref{fig:lse44_feige110} shows a comparison between the {\it
FUSE} spectra of LSE~44 and the sdOB star Feige~110 ($T_{\rm{eff}} = 42$,300~K
and $\log g = 5.95$; Friedman et al.  2002).  The figure illustrates that both
stars have many lines in common, but because of the lack of atomic data, we
cannot identify these absorption lines.  The \ion{Si}{4} lines at 1122.49~\AA\
and 1128.33~\AA\ and the \ion{P}{5} line at 1128.008~\AA\ are practically the
only stellar lines that can be identified in these portions of the {\it FUSE}
spectrum.  This example illustrates clearly that blends of unidentified stellar
lines with interstellar lines could lead to systematic errors when measuring
interstellar column densities.

Table 3 summarizes the important properties of LSE~44 and the sight line to
this star.

\section{Interstellar Column Density Analyses}

Two techniques have been used to estimate the column densities in this analysis.
The first involves directly measuring the equivalent widths of absorption lines,
which are fit to a single-component Gaussian curve-of-growth (COG) (Spitzer
1978).  Although it appears from the line widths that the line of sight may
traverse more than a single velocity component, no high resolution optical or UV
spectrum exists which would allow us to model a more complicated velocity
structure.  We therefore made the simplifying assumption that a single
interstellar cloud with a Maxwellian velocity distribution is responsible for
the absorption.  The estimates of the column density ($N$) and Doppler parameter
($b$) derived from the COG are likely to include systematic errors due to this
assumption.  For example, the derived $b-$value will not be a simple quadrature
combination of thermal and turbulent velocity components, as it would be for a
true single cloud.  Instead, it will also reflect the spread in velocities of
the multiple clouds that are likely to lie along a sight line of this length.
Jenkins (1986) has discussed the systematic errors associated with the COG
technique when there are multiple clouds of various strengths along a sight
line, and has shown that if many non-saturated lines are used, and if the
distribution of $N$ and $b-$values in the clouds is not strongly bimodal, then
the column density is likely to be underestimated by $\la$ 15\%.

Despite these shortcomings, the COG technique has the virtue of being unaffected
in principle by convolution with the instrumental line spread function (LSF).
In practice, it must be used with caution because of the problem of blending
with neighboring lines and, especially for a target like LSE 44, because it is
difficult to establish the proper continuum in some spectral regions, presumably
due to the presence of many unidentified stellar absorption features.

The second technique is profile fitting (PF).  For all species except \HI\ we
used the code {\tt Owens.f}, which has been applied to \fuse\ spectra for many
previous investigations.  This code models the observed absorption lines with
Voigt profiles using a $\chi^2$ minimization procedure with many free
parameters, including the line spread function, flux zero point, gas
temperature, and turbulent velocity within the cloud.  We split each spectrum
into a series of small sub-spectra centered on absorption lines, and fit them
all simultaneously.  Each fit typically includes about 60 spectral windows, and
approximately 100 transitions of \HI, \DI, \OI, \NI, and H$_2$ ($J=0$ to $5$).
The wavelengths and oscillator strengths of the lines used in the profile
fitting analysis are given in Table 4.  Note that there are fewer than 100
entries in the table because some lines appear in multiple detector segments.
\DI, \OI, and \NI\ were assumed to be in one component, \htwo\ in a second, and
\HI\ in a third.  Velocities between windows were allowed to shift to
accommodate inaccuracies in the \fuse\ wavelength calibration.  The line spread
function (LSF) is fixed within a spectral window, but is allowed to change in a
given detector segment from one window to another, and between segments.  This
is consistent with the performance of the \fuse\ spectrographs (Sahnow et al.
2000).  No systematic relationship between the LSF and the derived column
density or $b-$value of any species was observed.

Only one interstellar component for a given species was assumed along the sight
line although this is unlikely to be true for such a distant target.  H\'ebrard
et al.  (2002) discuss tests they performed on an extensive list of potential
systematic errors using profile fitting, including the single component
assumption.  In these tests they found that fits obtained with up to five
interstellar components gave the same total column density within the 1 $\sigma$
error bars, as long as saturated lines are excluded from the analysis, which we
have been careful to do.  Thus, we report total integrated column densities
along the sight line.

It is possible that the temperature and turbulent velocities differ from one
cloud to another.  To test this we did profile fits with \DI, \OI, and \NI\ in
three independent components.  The column density estimates differ by less than
1 $\sigma$ from those obtained by assuming a single component.  The $b-$values
also agree with that obtained by assuming a single component, although they are
not well-constrained because only unsaturated lines are used.  These species
could also have different ionization states if they reside in clouds with
different temperatures.  However, the ionization potentials of \DI\ and \OI\
and, to a slightly lesser extent, \NI\, are so similar that this is unlikely to
be a problem.

The \FeII\ absorption lines detected in the spectrum are abnormally broad for
interstellar lines, about 10 \kms, with a temperature of $\sim2\times10^5$ K,
compared to $\sim 6$ \kms\ for \NI\ and \OI\ (see \S4.2).  The cause of this
broadening is not known, but \FeII\ is not present in the stellar atmosphere
because the temperature is too high.  In the fits we did not include \FeII\ in
the component containing \DI, \OI, and \NI, but this has no effect on the
determination of the interstellar column densities of these species.

All laboratory wavelengths and oscillator strengths used in this work are from
Morton~(2003) and Abgrall et al.~(1993a, 1993b).  Neither the COG nor profile
fitting techniques include oscillator strength uncertainties in the error
estimates.  The application of {\tt Owens.f} to \fuse\ data is described in
greater detail by H\'ebrard et al.  (2002).

\subsection{The \DI\ Analysis}

We determined \NDI\ using only the profile fitting technique.  We attempted to
construct a COG, but we were unable to determine unambiguously the proper
placement of the continuum in the presence of the strong \HI\ lines separated
from the \DI\ lines by +81 \kms.  The situation is particularly difficult for
this sight line due to the unusual shape of the neighboring continuum around the
$\lambda 916.180$ line, and the unaccounted for absorption on the red wing of
the \HI\ $\lambda 920.712$ line.  Our tests indicated that failure to recognize
such problems caused COG estimates of \NDI\ to be approximately 0.3 dex too low.
Once the continuum is defined in the COG analysis, it is fixed.  By contrast,
although continuum placement errors will also lead to column density errors with
profile fitting, this technique allows the shape of the continuum to change as
part of the fitting process.  The continuum was modeled with polynomials of
order $0-4$ to allow for a large range of continuum fits.  We required
acceptable fits simultaneously to all five \DI\ lines used in the analysis (see
below).  In addition, removing any of the lines from the analysis did not change
the final column density estimate significantly.  This was not true for the COG
fit, which was highly sensitive to the continuum placement at the $\lambda
916.180$ line.  Removal of this line changed the COG result by more than 0.2
dex.  Therefore, by virtue of its simultaneous and self-consistent fit over
multiple lines, and its sensitivity to the exact way in which \DI\ blends with
the blue wing of the neighboring \HI\ line, profile fitting is the best method
for measuring the deuterium column density.

Five \DI\ absorption profiles over three separate spectral lines were used in
the profile fits:  $\lambda916.180$ in the SiC1B channel, and $\lambda919.100$
and $\lambda920.712$ in both the SiC1B and SiC2A channels.  Longer wavelength
\DI\ lines are too saturated to provide additional constraints on the column
density estimate, and the convergence of the \HI\ Lyman series precludes using
shorter wavelength lines.  The \DI\ $\lambda916.931$ and $\lambda917.880$ lines
are too strongly blended with \OI\ and \htwo\ lines to be used.  Although the
same transition observed in separate \fuse\ channels potentially are subject to
systematic errors arising from similar effects, such as continuum placement and
blending, they do represent independent measurements with distinct
spectrographs.  Thus, we include all five profiles in our analysis.

To properly measure the interstellar \DI, the effects of the stellar \HI\ Lyman
series and \ion{He}{2} absorption must be removed.  This was done in two ways.
First, the stellar model described in \S3 was shifted in velocity space and
scaled in flux until the stellar \HI\ damping wings matched the observed
spectrum away from the cores of the interstellar \HI\ lines.  The resulting
average velocity difference was $\Delta V \equiv V_* -V_{\rm ISM} = -39\pm5$
\kms\ and $-32\pm13$ \kms\ for the SiC1B and SiC2A spectra, respectively.  This
compares favorably to $\Delta V = -33\pm6$ \kms\ based on the measured
velocities in the high-dispersion \iue\ spectrum of photospheric lines
(\ion{N}{4}, \ion{N}{5}, \ion{O}{6}, \ion{Si}{6}, \ion{S}{5}, \ion{C}{3}, and
\ion{C}{4}) and interstellar lines (\ion{Si}{2}, \ion{Si}{3}, and \ion{S}{2}).
The spectral resolution of \iue\ in this mode is approximately 25 \kms\ (FWHM).

The stellar model is only an approximation of the true absorption from the
atmosphere of LSE 44.  Since not all species present in the atmosphere are
properly accounted for in the model, and since some of the atomic constants
included in the model are not well-known, using the model may in fact introduce
systematic errors.  To check for this we also applied the profile fitting
without the model, using instead up to a fourth-order polynomial to fit the
stellar spectrum and continuum.  This method has been used many times before in
similar analyses; see, for example, H\'ebrard \& Moos  (2003).  As described
below, the two methods give column densities which agree within the errors.

\begin{figure}
\epsscale{0.6}
\plotone{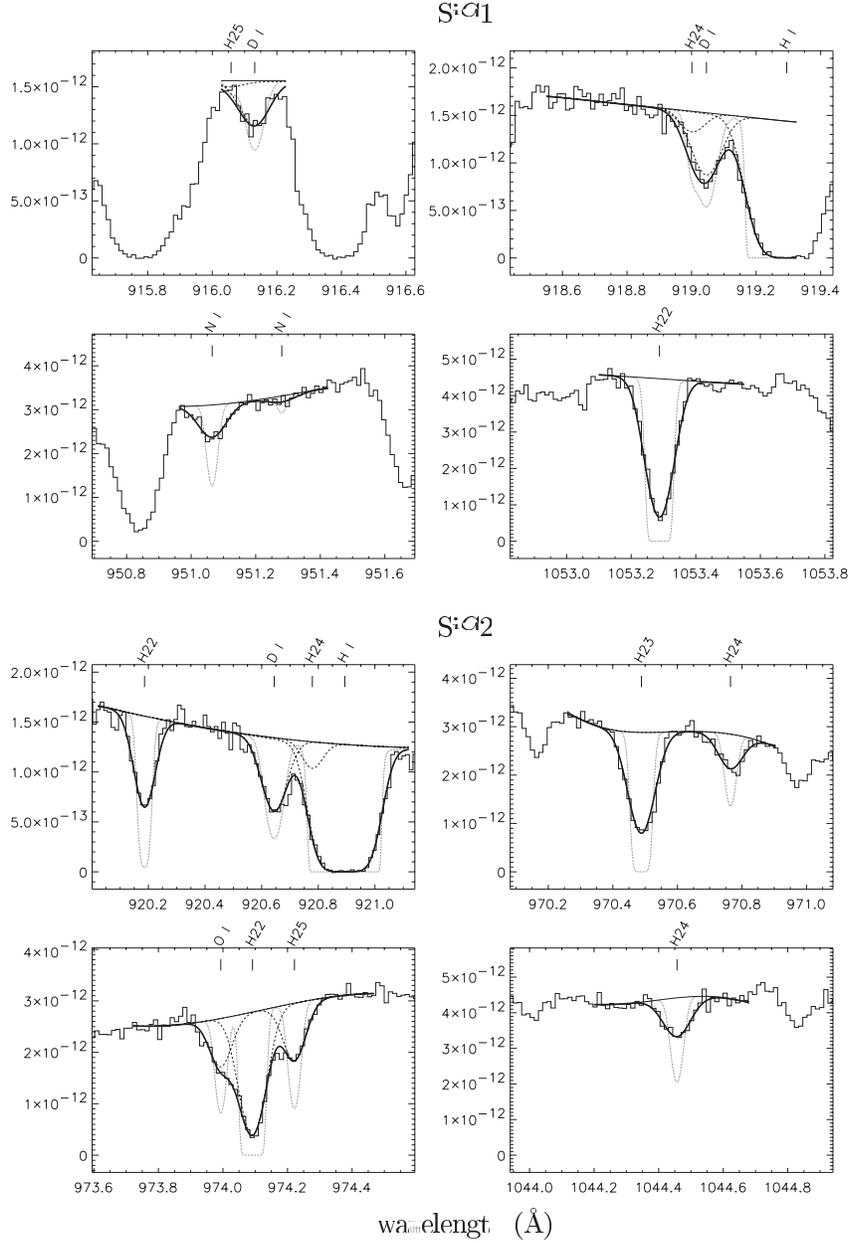}
\caption{Examples of spectral windows used in the profile fitting analysis.
Histogram lines are the data, heavy black lines are the fits, light black lines
are the continua, and the dashed lines are the fits for the individual species.
The dotted lines are the model profiles prior to convolution with the LSF.  The
\htwo\ lines of levels $J=2$ and $J=5$ are designated as H22 and H25.  The
X-axis is wavelength in \AA\ and the
Y-axis is flux in units of ergs cm$^{-2}$ s$^{-1}$ \AA$^{-1}.$ Eight
spectral windows are displayed here, including 18 transitions.  However, the
complete fit includes typically 60 spectral windows in all \fuse\ detector
segments, and approximately 100 transitions.
\label{fig_owens}}
\end{figure}

Several parameters are free to vary through the fits, including the column
densities, the radial velocities of the interstellar clouds, and the shapes of
the stellar continua.  Some instrumental parameters are also free to vary,
including the widths of the Gaussian line spread function (LSF), which is
convolved with the Voigt profiles within each spectral window.  The averaged
width of the LSF that we found is 5.8 pixels, with a 0.8-pixel $1\,\sigma$
dispersion (full widths at half maximum).  These {\it FUSE} pixels are
associated with CALFUSE version 3, and are 0.013\AA\ in size.  Note that CALFUSE
versions 1 and 2, which was used in most previously published \fuse\ studies,
produced pixels approximately half this size.  This explains why the LSF
reported previously was approximately about 11 pixels wide (see, e.g., Friedman
et al.~2002; H\'ebrard et al.~2002).  Some examples of fits are plotted in
Figure~\ref{fig_owens}.

Using the stellar model we find log $N$(\DI)$_{sm} = 15.84\pm{0.08}$.  Using a
polynomial fit to the stellar plus interstellar continuum, we find
$N$(\DI)$_{poly} = 15.89\pm{0.08}$.  These errors include our estimates of the
systematic errors associated with the stellar normalization.  Our final estimate
is the mean of these two values\footnote{Unless otherwise noted, errors on all
quantities in this paper are $2\,\sigma.$}, log \NDI = $15.87\pm{0.08},$ where
the error of the mean has not been reduced below the individual values since
$N$(\DI)$_{sm}$ and $N$(\DI)$_{poly}$ are not independent estimates of the
column density.

\subsection{The \OI\ Analysis}

The continuum placement problem is much less severe for the \OI\ lines because
the spectral regions adjacent to the lines used in the analysis are generally
smooth and well-behaved, without the presence of very strong adjacent lines.
Thus, we use both the profile fitting and curve-of-growth techniques to estimate
\NOI.  The profile fitting analysis used only the \OI\ $\lambda 974.070$ line,
because this line is almost completely unsaturated.  We did verify that the fit
to other \OI\ lines in the spectrum was acceptable, but they provide almost no
additional constraint on the column density because they are saturated.  One
might be concerned that an erroneous $f-$value for this transition would cause
an inaccurate estimate of \NOI.  However, H\'ebrard et al.  (2005) discuss this
issue, and show that it is consistent with column densities derived from
stronger \OI\ lines in the case of BD+28$\arcdeg4211,$ and derived from the weak
$\lambda1356$ intersystem line in the case of HD 195965.  The \OI\
$\lambda974.070$ line is blended with two \htwo\ lines, Lyman 11-0R(2)
$\lambda974.156$ and Werner 2-0Q(5) $\lambda974.287,$ and absorption from these
species was accounted for when calculating \NOI, as described below.  There is
also an \htwo\ $J=6$ line at nearly the same wavelength, but the column density
in this rotational level is too small to be of concern.  The result of our
analysis is $N$(\OI)$_{pf} = 17.53\err{0.25}{0.15}.$

To construct the curve-of-growth the equivalent widths of the \OI\ lines were
measured in the following way.  First, low-order Legendre polynomials were
fitted to the local continua around each \OI\ line.  The lines were integrated
over velocity limits chosen to exclude absorption from adjacent lines.  Only
isolated lines were selected so that no deblending was required.  Scattered
light remaining after the CALFUSE 3 reduction is negligibly small, and no
additional correction was required.  The measured equivalent widths are given in
Table 5.  The estimated errors from this method are described in detail by
Sembach \& Savage (1992).  They include contributions from both statistical and
fixed-pattern noise in the local continuum.  Continuum placement is particularly
difficult and subject to error due to the presense of many unidentified stellar
lines.  To estimate the magnitude of such errors, we did trials in which we drew
the continuum at its maximum and minimum plausible locations, and compared the
calculated equivalent widths with the best estimated placement.  We added 2
m\AA\ to the error of each measured \OI\ line equivalent width to account for
these placement errors.  This is consistent with the magnitude of this
systematic error determined in previous analyses of similar sight lines (e.g.,
Friedman et al., 2002).

\begin{figure}
\epsscale{0.90}
\plotone{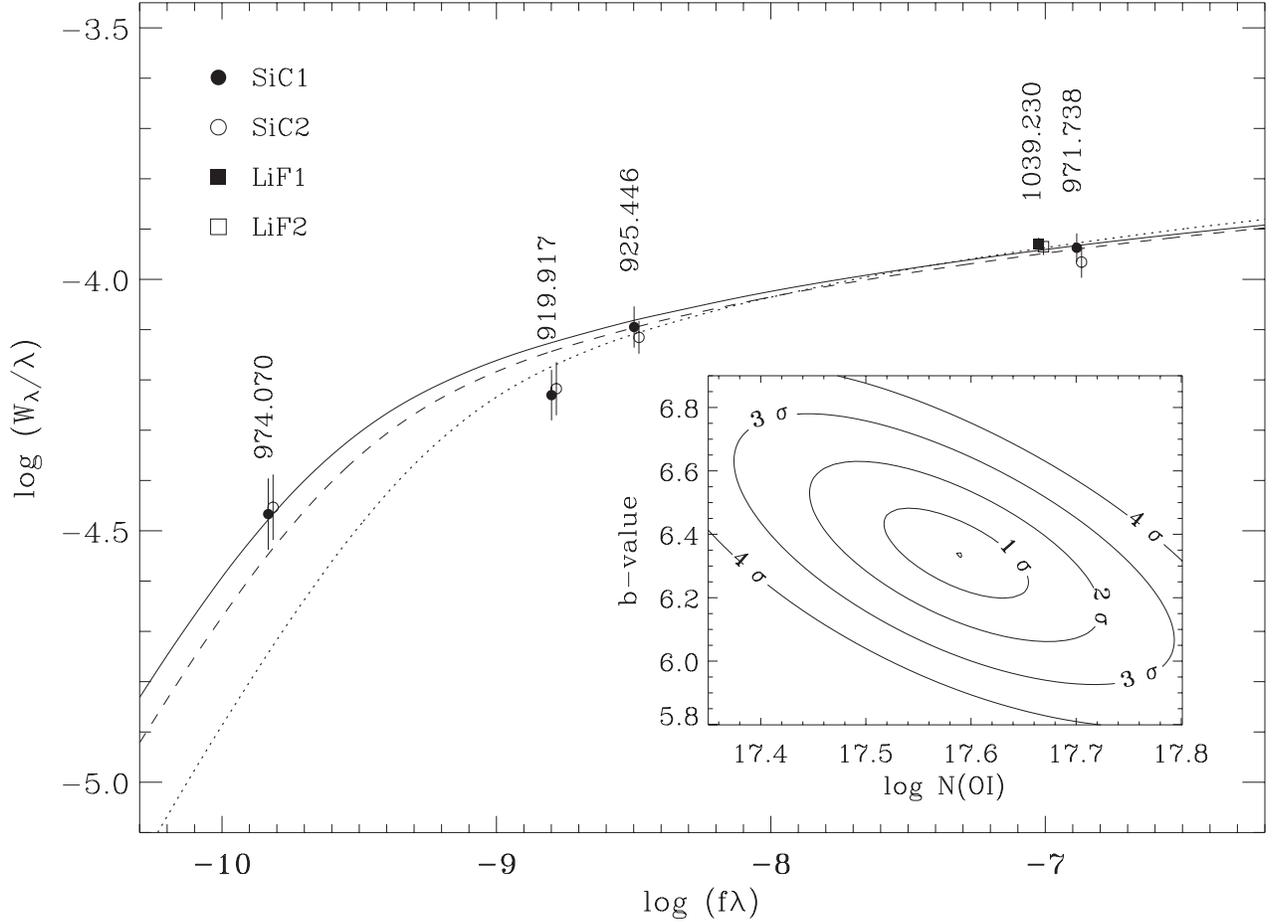}
\caption{The single component, Gaussian curve of growth for \OI.  The best fit
solution, which excludes the $\lambda919$ line, is log $N$(\OI) =
$17.59\err{0.13}{0.14}$ and $b = 6.33\err{0.30}{0.26}$ (solid line).  Including
all lines gives log $N$(\OI) = $17.52\err{0.22}{0.23}$ and $b =
6.33\err{0.49}{0.44}$ (dashed line).  The solution if both the $\lambda 974$ and
$\lambda919$ lines are excluded is log $N$(\OI) = $17.22\err{0.55}{0.29}$ and $b
= 6.85\err{0.57}{0.75}$ (dotted line), which demonstrates that the lack of
optically thin lines can result in a large error in the column density estimate.
See \S4.2 for additional discussion.  For clarity, at each wavelength the data
points derived from each of the \fuse\ channels have been slightly separated.
The inset shows the log($N$)/$b-$value error contours for the best fit solution.
\label{fig_OI_cog}}
\end{figure}

Due to the blending with the \htwo\ lines we initially did not include the \OI\
$\lambda974.070$ line.  However, this line is extremely important because it is
the only \OI\ line that is not substantially saturated, and therefore provides
the greatest constraint on \NOI.  We constructed curves of growth for the $J=2$
and $J=5$ rotational levels of \htwo, calculated column densities and
$b-$values, determined the appropriate Voigt profiles, and used them to remove
the \htwo\ signature from the blended line.  Then the equivalent width of the
\OI\ line was measured in the usual way.  The \htwo\ analysis is described in
Section 4.5.  The \OI\ curve of growth is shown as the solid line in
Figure~\ref{fig_OI_cog}.  The best fit column density is log $N$(\OI)$_{COG} =
17.59\err{0.13}{0.14}$.  Combining this with our profile fitting result, we
arrive at our best estimate of the \OI\ column density, log \NOI =
$17.57\err{0.21}{0.15}.$

We note that the \OI\ $\lambda919.917$ line has been excluded from the analysis.
We were unable to get a good fit of this line with profile fitting, and it fell
well below the curve-of-growth.  This is also true for Feige 110 (H\'ebrard et
al.  2005), whose spectrum is similar in many respects to that of LSE 44, as
discussed in \S3.  It is unlikely that this line has a significantly incorrect
$f-$value since the absorption from this line is consistent with absorption from
other \OI\ lines in, for example, the spectrum of BD+$28\arcdeg4211$ (H\'ebrard
\& Moos 2003).  It is possible that there is unidentified absorption adjacent to
this line, which depresses the local continuum, causing an underestimate of the
equivalent width of this line.  Alternatively, the location of this line on the
COG may be an indication that there are multiple components along the sight
line, with different column densities.  This would cause a "kink" in the COG,
which might be more obvious if additional \OI\ lines were available to include
on the plot.  However, with no high resolution spectra available, we are unable
to confirm this hypothesis.  If we include the $\lambda919.917$ line, the
single-component COG gives log \NOI = $17.52\err{0.22}{0.23}$ and $b =
6.33\err{0.49}{0.44},$ which is shown as the dashed line in
Figure~\ref{fig_OI_cog}.  This differs from our best estimate column density by
less than $1\sigma.$

It is worth emphasizing the importance of non-saturated lines in the
determination of \NOI\, even at the expense of the potential introduction of
systematic errors associated with using strongly blended lines.  Computing a COG
excluding the \OI\ $\lambda974.070$ (and $\lambda919.917$) lines gives $N$(\OI)
= $17.22\err{0.55}{0.29}$, about $2\sigma$ less than the result obtained when
including the weak line.  This is shown as the dotted line in
Figure~\ref{fig_OI_cog}.  The fit to the remaining spectral lines is quite
acceptable, but is clearly inconsistent with the weak line.  Indeed, this line
was not used in the initial analysis of Feige 110 (Friedman et al.  2002).  We
have recently included $\lambda974.070$ in a re-analysis of \NOI\ along this
sight line.  The COG and profile fitting methods are now in good agreement, and
the column density estimate has increased by 0.33 dex compared to the original
estimate.  This is discussed more fully in H\'ebrard et al.  (2005).

\subsection{The \NI\ Analysis}

As was the case with \DI, only profile fitting was used to determine \NNI.  The
COG technique was not used because of the uncertainties in establishing
continuum levels on each side of the \NI\ lines.

Three lines were used in the profile fitting analysis:  $\lambda 951.079,
\lambda 951.295,$ and $\lambda 955.882$.  The method is the same as was used for
the deuterium analysis, and is described in \S4.1.  The result is log $N$(\NI) =
$16.43\pm0.14.$

\subsection{The \HI\ Analysis}

In the direction of LSE 44, the \lya\ line provides the best constraint on the
total \HI\ column density because its radiation damping wings are very strong.
However, \lya\ is not accessible to \fuse.  \lyb\ displays modest damping wings,
and we attempted to model the absorption profile in order to estimate \NHI.  We
were unable to do this primarily because of the presense of two \htwo\ lines,
J=0 $\lambda1024.230$ and J=2 $\lambda1026.526$, on the blue and red wings of
the \lyb\ absorption, respectively.  As discussed in \S4.5, all lines of \htwo\
in the J=0 rotational level lie on the flat portion of the COG, so we are unable
to make even a rough estimate of the column density.  The situation for J=2 is
only slightly better, with the weakest lines still considerably saturated.
Since these lines are located almost exactly at the regions of the \lyb\ profile
which are most sensitive to constraining a model profile, it is impossible to
obtain a good estimate of \NHI.  The higher order Lyman lines in the \fuse\
spectrum are even weaker relative to the stellar absorption than is \lya\ or
\lyb.  Correcting these lines for stellar absorption would therefore require
more precise knowledge of the proper flux scaling of the model, and of the
velocity difference between the stellar and interstellar components.  We have
examined the range of $b-$value/column density combinations allowed by the
higher Lyman series lines in the \fuse\ spectrum, along with the uncertainties
in correcting for the stellar absorption, and we find that these lines do not
constrain $N$(\HI) with sufficient precision to provide an interesting D/H
measurement.  No suitable {\it HST} spectra of the LSE 44 \lya\ line have been
obtained.  Consequently, we have used the only high-dispersion echelle
observation of this star obtained with the short wavelength prime camera on
\iue\ (exposure ID SWP29086) to estimate $N$(\HI) toward LSE 44.  This \iue\
mode provides a resolution of $\sim$25 \kms\ FWHM and covers the 1150-1950 \AA\
range.

The \iue\ observation was obtained on 1986 August 29 with an exposure time of
16.5 ksec.  We retrieved the \iue\ spectrum from the STScI MAST archive with
both IUESIPS and NEWSIPS processing.  Since there are several known problems
with NEWSIPS processing of high-dispersion observations (Massa et al.  1998)
that can adversely affect the $N$(\HI) measurement, particularly the background
subtraction in the vicinity of \lya\ where the orders are closely spaced on the
detector (Smith 1999), we generally prefer IUESIPS spectra.  Nevertheless, we
compared the IUESIPS and NEWSIPS versions of the spectrum at the outset.  In
addition to the standard processing, we applied the Bianchi \& Bohlin (1984)
correction method to the IUESIPS spectrum to compensate for scattered light from
adjacent orders in the background regions near the interstellar \lya\ line.

\subsubsection{Analysis Method}

To measure the total interstellar \HI\ column density, we used the method of
Jenkins (1971); see also Jenkins et al.  (1999) and Sonneborn et al.  (2000).
In brief, we constrained the \HI\ column density using the Lorentzian wings of
the \lya\ profile, which have optical depth $\tau$ at wavelength $\lambda$ given
by $\tau (\lambda ) = N$(\HI)$\sigma (\lambda ) = 4.26 \times 10^{-20}
N$(\HI)$(\lambda - \lambda _{0})^{-2}$.  Here $\lambda _{0}$ is the centroid of
the interstellar \HI\ absorption in this direction, which was determined from
the \ion{N}{1} triplet at 1200 \AA .  We estimated the value of $N$(\HI) that
provides the best fit to the \lya\ profile by minimizing $\chi ^{2}$ using
Powell's method with five free parameters:  (1) $N$(\HI), (2)-(4) three
coefficients that fit a second-order polynomial to the continuum (with a model
stellar \lya\ line superimposed, see below), and (5) a correction for the flux
zero level.  We then increased (or decreased) $N$(\HI) while allowing the other
free parameters to vary in order to set upper and lower confidence limits based
on the ensuing changes in $\chi ^{2}$.

In addition to the \lya\ absorption profile due to interstellar \HI\ , a
subdwarf O star like LSE 44 will have a substantial stellar \lya\ absorption
line with broad, Lorentzian wings as well.  Neglect of this stellar line when
setting the continuum for fitting the interstellar \lya\ will lead to a
substantial systematic overestimate of $N$(\HI).  We have used the stellar
atmosphere models discussed in \S 3 to account for the stellar \lya\ line.  The
offset of the stellar line centroid with respect to the interstellar lines was
determined by comparing the velocities of several stellar and interstellar lines
in the \fuse\ spectrum, and the uncertainty in the determined offset has a
negligible impact on the derived $N$(\HI).  We use the most shallow and deepest
stellar profiles derived from a grid of models, as described in \S 3, to set
upper and lower confidence limits on $N$(\HI) in the following section.

\subsubsection{$N$(\HI) Results}

\begin{figure}
\epsscale{0.90}
\plotone{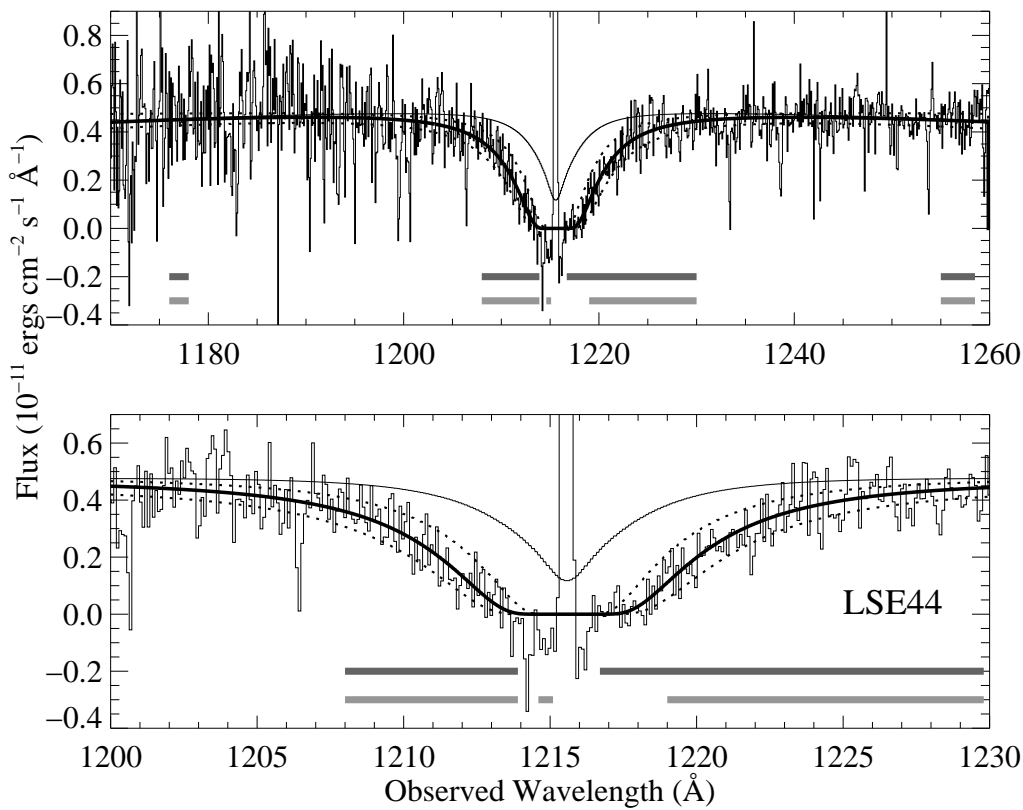}
\caption{High-dispersion {\it IUE} observation of LSE 44 (histogram) with the
best fit \ion{H}{1} profile (black line through spectrum) and upper and lower
bounds (dotted lines) on $N$(\ion{H}{1}) at the $2\,\sigma$ confidence level.
The highest curve is the stellar model, described in \S3.  We have used the
spectral region indicated by the dark horizontal line in the fit.  Note that
this uses the region from approximately $1216.5-1218.1$\AA, in the core of the
saturated line, to establish the flux zero point.  We initially used the
spectral region indicated by the gray horizontal line for the fit, but
determined that this incorrectly set the zero point of the flux.  See \S4.4.2
for additional discussion.  The upper and lower panels plot the same data and
fits on different scales to enable the reader to inspect the details in the
\lya\ profile as well as the continuum fit well away from the interstellar
\ion{H}{1} line.  The emission line in the center of the saturated Ly$\alpha$
core is geocoronal Ly$\alpha$ emission, and is excluded from the analysis of the
interstellar \HI\ column density.
\label{fig_hifit}}
\end{figure}

Figure~\ref{fig_hifit} shows the \iue\ spectrum of LSE 44; both panels plot the
same data but the upper panel shows a broader wavelength range to enable the
reader to inspect the fit to the continuum away from the strong \lya\
absorption.  Overplotted on the spectrum with dotted lines are the fits that
provide $2\,\sigma$ upper and lower limits on the interstellar $N$(\HI).  The
deepest model stellar \lya\ line is assumed for the lower limit on $N$(\HI), and
the most shallow stellar \lya\ model is assumed for the upper limit.  From these
fits we derive log $N$(\HI) = 20.52$^{+0.20}_{-0.36}$ ($2\,\sigma$).

There are, of course, many potential sources of systematic error in the $N$(\HI)
measurement.  We have already discussed the stellar \lya\ line.  Strong, but
nevertheless unrecognized, narrow stellar lines within the \lya\ profile are
another potential source of systematic error.  We have included stellar lines in
the continuum regions used to estimate $\sigma$ for the $\chi ^{2}$ calculation,
but nevertheless future higher resolution and signal-to-noise observations that
resolve strong stellar lines might yield a substantially different result.
However, it seems likely that the greatest source of systematic errors in this
case is location of the zero-point flux.  In a saturated line the core should be
broad and completely black.  However, examination of Figure \ref{fig_hifit}
shows that the core has significant undulations.  We have decided that the
flattening of the spectrum between approximately 1216.5\AA\ and 1217.8\AA\ gives
the best indication of the zero point.  The spectral regions used in the fit to
determine \NHI\ are indicated in Figure~\ref{fig_hifit} by the dark horizontal
lines below the spectrum.

We recognize that there are other plausible zero-point placements.  For example,
Friedman et al. (2005) initially used the spectral region indicated by the
gray horizontal line in Figure~\ref{fig_hifit} to determine the zero point,
completely excluding the flat part in the red side of the airglow emission.  The
result of this fit was log $N$(\HI) = 20.41$^{+0.19}_{-0.33}.$ However, we
ultimately decided that the flattening to the red of the airglow line better
represents the expected shape of the line core, and the undulations on the blue
side make this region an unreliable indicator of the zero point.  The undulations
could have many sources, including improper joining of spectral orders in the
\iue\ echellogram.  We note that if the region on the blue side is used to set
the zero point, we obtain a {\it lower} value for \NHI, which in turn
makes D/H even higher than the value we obtain in the next section.

\subsection{The \htwo\ Analysis}

\begin{figure}
\epsscale{0.90}
\plotone{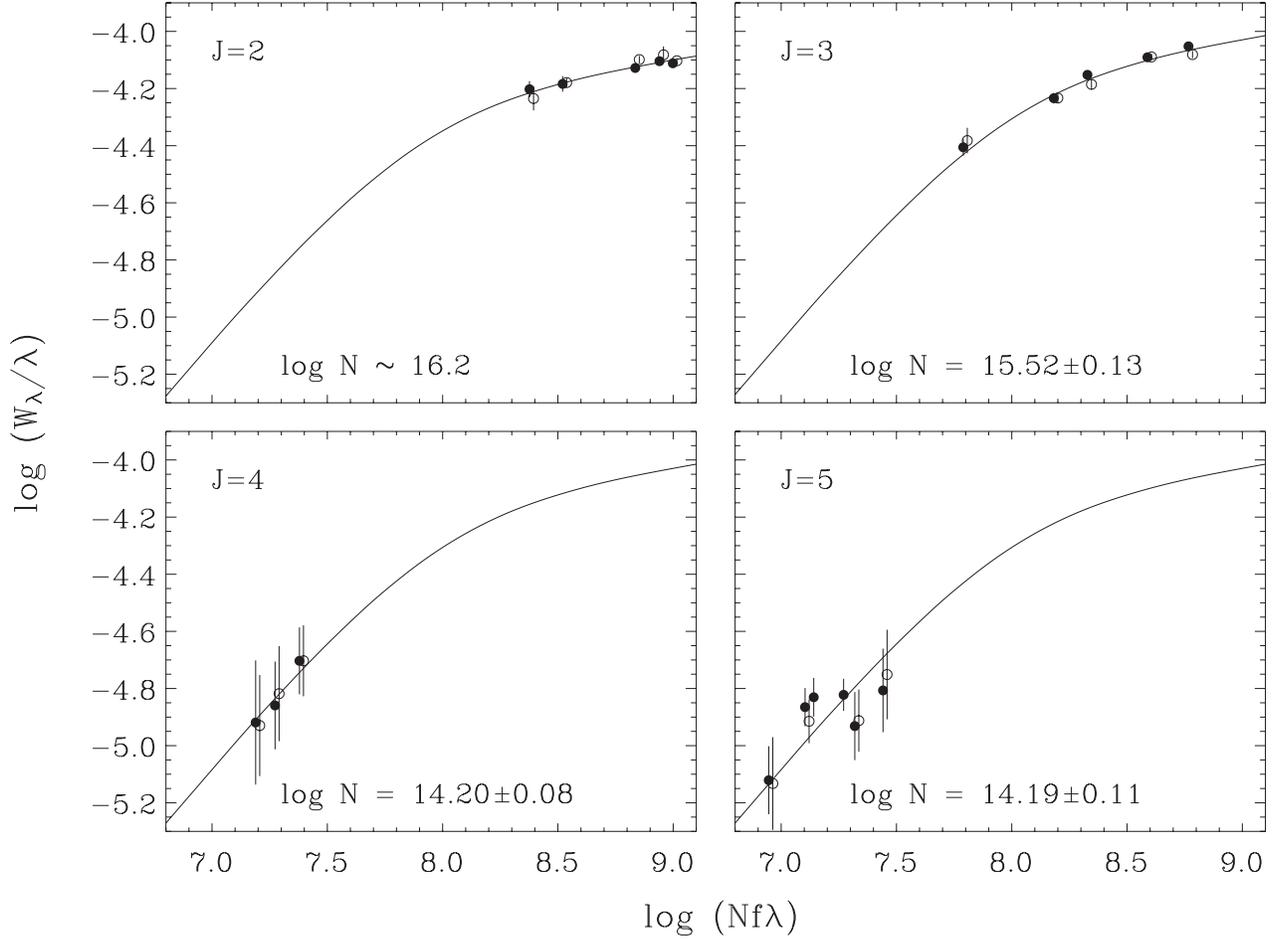}
\caption{Curves-of-growth for \htwo\ $J=2-5$ rotational levels.  All curves are
ploted over the same range of log (Nf$\lambda$) to clearly display the
rotational levels for which $N(H_2)$ is well-constrained.  For $J=2$ it is not
possible to detemine a meaningful column density error because all lines are
saturated.  The $b-$value is well constrained only for $J=3$, for which
$b=7.54\err{0.53}{0.44}$ \kms.  We set $b=7.54$ when computing the COG for $J=4$
and $J=5$, although the column densities are insensitive to the choice of $b$. 
Filled symbols are data from SiC1B channel, and open symbols from SiC2A.
\label{fig_h2cog}}
\end{figure}

Accurate measurements of the column densities of molecular hydrogen in various
rotational levels are important so that the \htwo\ lines may be deblended in
order to measure the column densities of \DI\ and metals.  For example, as noted
in \S4.2, the weak \OI\ $\lambda974.070$ line is blended with $J=2$ and $J=5$
lines.  We attempted to measure $N$(\htwo) for $J=0-5$ so that we could estimate
the total \htwo\ column density and determine the molecular fraction of the gas
along the sight line.  However, the $J=0$ and $J=1$ rotational levels could not
be accurately determined.  In each case there are no lines strong enough to show
well-developed damping wings, and no lines weak enough to be on the linear
portion of the curve-of-growth.  The situation is slightly better for $J=2,$
which has some weaker lines but none that are in the true linear regime.  We
constructed a COG, but the column density is not well-constrained.  For $J=3-5,$
weaker lines are available and the errors are lower. The measured equivalent
widths are given in Table 6, and the column densities are given in Table 7.
The \htwo\ COGs are shown in Figure~\ref{fig_h2cog}.

\section{Results and Discussion}

\begin{figure}
\epsscale{0.90}
\plotone{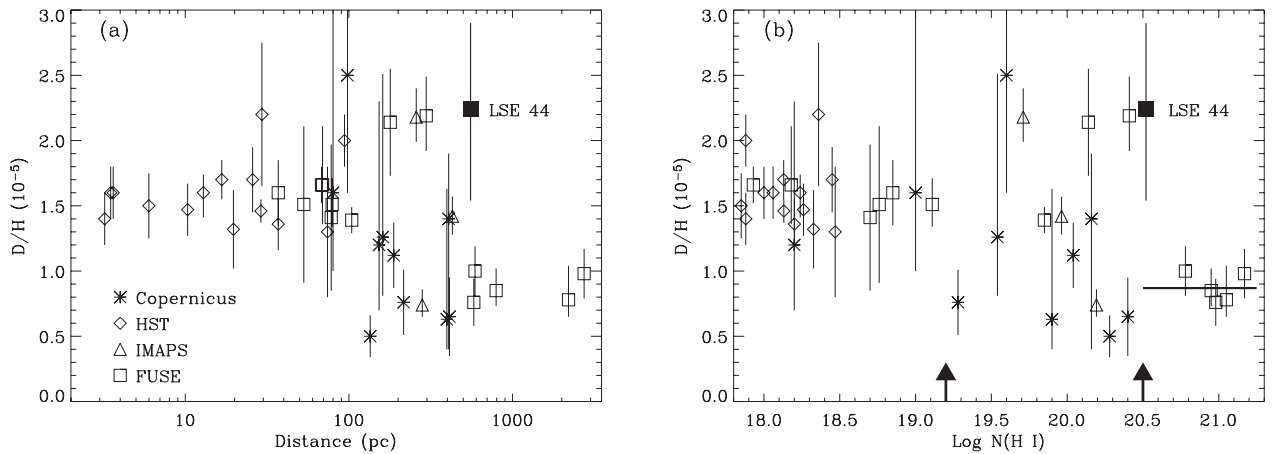}
\caption{D/H as a function of (a) distance and (b) \HI\ column density.  In this
figure, adapted from Wood et al.  (2004), the $1\sigma$ errors are plotted.  The
filled square shows the data point from this study.  The arrows in panel (b)
indicate the Wood at al.  boundary of the Local Bubble, at log \NHI\ = 19.2, and
the start of the third D/H regime, at log \NHI\ = 20.5, where D/H is averaged
over many regions, leading to a roughly constant value.  The horizontal lines
indicate the average values of D/H for log \NHI $< 19.2$ and log \NHI $> 20.5$
(excluding LSE 44), respectively. Mixing caused by supernovae may be
responsible for the uniform D/H distribution within $\sim100$ pc. Also included
in these plots are results from Williger et al. (2005) and H\'ebrard et al.
(2005).
\label{fig_dhplot}}
\end{figure}

We have measured the column densities of \DI, \OI, \NI, \HI, and \htwo\ along
the sight line to LSE 44.  This target is of particular interest because it is
one of only five for which such values are known for Galactic targets at
distances greater than about 500 pc and hydrogen column densities log \NHI $>
20.5.$ Table 8 gives a summary of the results of this study, and
Figure~\ref{fig_dhplot} shows our D/H value compared to previous measurements.
Our principle results are D/H = $(2.24\err{1.39}{1.32}) \times 10^{-5},$ D/O =
$(1.99\err{1.30}{0.67}) \times 10^{-2}$, and O/H = $(1.13\err{0.96}{0.71})
\times 10^{-3}$.  Of the targets with the \HI\ column densities only LSE 44 and
PG0038+199 (Williger et al.  2005) have high D/H, about $2.2 \times 10^{-5}.$
D/H for the remaining targets are all $\leq 1.0 \times 10^{-5}.$

The spatial variation of D/H has drawn increased attention in recent years.  The
large number of measurements made using \fuse, together with earlier results
from \imaps, \hst, and \copernicus, led to initial speculation that the observed
variations in D/H and O/H were due to differing levels of astration (Moos et al.
2002).  Models of galactic chemical evolution (de Avillez \& Mac Low 2002;
Chiappini et al.  2002) attempt to account for the spatial variability in terms
of mixing scale lengths and times.  One of the greatest difficulties for such
models is to properly account for the astration factor required to reduce the
primordial deuterium abundance to that observed along some sight lines in the
Galaxy.  The primordial value, estimated from gas-phase measurements of
low-metallicity material at high redshift toward 5 quasars, is (D/H)$_{prim} =
2.78\err{0.44}{0.38} \times 10^{-5}$ (Kirkman et al.  2003).  This is consistent
with the value inferred from the amplitude of the acoustic peaks in cosmic
background radiation (CBR) measurements made with \wmap, (D/H)$_{CBR} =
2.37\err{0.19}{0.21} \times 10^{-5}$ (Spergel et al.  2003).  Thus, astration
factors of approximately 1.6 or 1.9 are required to account for Local Bubble D/H
abundances determined by Wood et al.  [(D/H)$_{LB} = (1.56\pm{0.04}) \times
10^{-5}$], and H\'ebrard \& Moos [(D/H)$_{LB} = (1.32\pm{0.08}) \times
10^{-5}$], respectively.  An astration factor of approximately 3 is required to
account for the mean D/H value of the four highest column density sight lines in
Figure~\ref{fig_dhplot}(b), and this is pushing the limits of most chemical
evolution models (Tosi et al.  1998).  These models will also have to account
for the deuterium abundance measured in the Complex C, a low-metallicity,
high-velocity cloud falling in to the Milky Way.  Sembach et al.  (2004) found
(D/H)$_{complex\,C} = (2.2 \pm 0.7) \times 10^{-5}$ and (D/O)$_{complex\,C} =
0.28 \pm 0.12\, (1\sigma).$

H\'ebrard \& Moos (2003) suggest that the low D/H value at large distances truly
reflects the present-epoch D/H abundance.  This conclusion is primarily based on
D/O and O/H measurements.  If this is correct, it presents a challenge to models
of galactic chemical evolution, which not only have to account for high
astration values, but also the observed spatial variation of these quantities.

Jura (1982) first proposed that deuterium might be depleted onto dust grains in
the interstellar medium.  Draine (2004a, 2004b) extended this by suggesting that
D/H in carbonaceous dust grains can exceed the overall D/H ratio by a factor
greater than $10^4$, if the material is in thermodynamic equilibrium.  Thus, it
is possible that a significant reservoir of deuterium is depleted onto these
grains, as long as the metallicity is approximately solar or greater.
Prochaska, Tripp, \& Howk (2005) have tested this idea by examining the
correlation between D/H and the highly depleted ion \TiII.  They find these are
correlated at greater than 95\% confidence level.  They emphasize, however, that
since not all data points support this relationship, measurements of distant
targets, such as the one in the current study, are most needed to further
constrain this relationship.  At present, we are unable to comment directly on
this hypothesis because the \TiII\ abundance along this sight line is not known.

Wood et al.  (2004) and Linsky et al.  (2005, in preparation) have applied Draine's suggestion
to the observational data, and suggest the presence of three D/H regimes, as
shown in Figure~\ref{fig_dhplot}.  In the Local Bubble, where log \NHI $\la
19.2,$ most of the deuterium is in the gas phase, presumably because of the
recent passage of strong shocks from supernovae or stellar winds.  Thus,
(D/H)$_{LBg} = (1.56 \pm 0.04) \times 10^{-5}$ is relatively high, spatially
uniform, and approximates the true local Galactic disk value.  At great
distances, where log \NHI $\ga 20.5$, sight lines traverse many regions, some
recently shocked and others not.  D/H is likely to be indicative of an average
of these regions.  However, the highest column density sight lines, which tend
to be the lengthiest ones, will be weighted toward the densest, coldest regions,
and presumably the most heavily deuterium-depleted clouds.  This may explain the
low D/H values toward the majority of the most distant targets.  In the
intermediate column density regime, 19.2 $\la$ log(\NHI) $\la$ 20.5, sight lines
traverse fewer regions with different shock event histories and therefore
exhibit more variability.

H\'ebrard \& Moos (2003) show that D/O tends to be uniform within the Local
Bubble, with a mean of $(3.84\pm 0.16) \times 10^{-2} (1\sigma),$ and variable
beyond it.  D/N is similar, but exhibits somewhat more variability due to
ionization effects.  At the highest column densities D/O and D/N converge to
mean values that are lower than the Local Bubble values.  Indeed, for all seven
sight lines for which log \NHI\ $> 20.1$ and log \NDI\ $>15.4$ (Hoopes, et al.
2003; Wood et al.  2004; Williger et al.),
including LSE 44, D/O lies in the relatively narrow range $(1.8 - 2.6) \times
10^{-2}.$ The fact that D/O is much more uniform than D/H is a surprise,
and suggests to some investigators that measurements of \NHI\ may suffer from
additional unidentified errors (H\'ebrard et al. 2005).

The $\sim100$ pc length scale in the Local Bubble over which D/H is roughly
constant (Figure~\ref{fig_dhplot}) may be a natural consequence of
supernova-driven mixing within the ISM.  Moos et al.  (2002) discuss a simple
model which has two time scales:  the mixing time and the time between
supernovae (SNe).  In a fully ionized ISM at $\sim10^6$K, the adiabatic sound
velocity is $c_s \approx 150$ \kms.  For a spherical parcel of gas with radius
100 pc, the crossing time is $t_s \approx 7 \times 10^5$ years.  Other
mechanisms, such as the Alfv\'en crossing time for a cloud with a magnetic
field, are generally slower.

A supernova rate of 0.02 yr$^{-1}$ (Cappellaro et al.  1993) spread uniformly
over the galactic disk of radius 8.5 kpc, yields approximately $8.8 \times
10^{-11}$ SNe pc$^{-2}$ yr$^{-1}.$ In the 100 pc radius parcel of gas considered
above, there would be one supernova every $4 \times 10^5$ yr.  Therefore, in
this simple model, $t_{SN} \sim t_s.$ Thus, one might expect that within the
Local Bubble, the material is well-mixed, and D/H, O/H, and N/H will be
approximately constant, with relatively small target-to-target variations.
Toward targets at greater distances, different supernovae events may have given
rise to interstellar regions with different enrichment levels, and these do not
have time to mix thoroughly with each other.  See Moos et al.  (2002) for
additional discussion of this model.  More sophisticated treatments of mixing,
including the effects of supernovae and infall of material into the Galactic
disk, are given by de Avillez (2000), de Avillez \& Mac Low (2002), and
Chiappini et al.  (2002).

If this model is correct, it is hard to explain the uniformity of D/O for the
high column density sight lines, unless a different mechanism forces this to be
the true Galactic value.  In addition, for these targets D/H and O/H vary by
factors of 2.9 and 4.6, respectively, but there is no convincing anticorrelation
between them, as might be expected if O is produced as D is destroyed in stellar
interiors.  Steigman (2003) reported a tentative detection of such an
anticorrelation based on \imaps\ and the initial set of \fuse\ results, all of
which had column densities lower than those considered here.

Neither the depletion hypothesis nor the low Galactic D/H hypothesis appears to
be consistent with all of the data currently available.  H\'ebrard et al.
(2005) show that the dispersion of D/H is greater than that of D/O in the sense
that there are no high D/O values at large distances or large column densities.
They argue against the depletion hypothesis since for sight lines beyond the
Local Bubble, the uniformity of D/O and O/H (Meyer, Jura, \& Cardelli 1998;
Meyer 2001; Andr\'e et al.  2003) strongly imply that D/H must be uniform as
well, unless D/O and O/H are correlated in exactly the same way.  They emphasize
that measurements of D/H may be subject to fewer systematic errors then
measurments of D/O and O/H.  On the other hand, nobody has identified an error
that could account for the D/H variability shown in Fig.~\ref{fig_dhplot}.  Two
observational tests will help clarify these interpretations.  First, additional
measurements of the depletion of refractory species, such as Fe, Si, or Ti
(Prochaska, Tripp, \& Howk 2005) along these sight lines would reveal whether a
correlation exists between depletion and the gas phase abundance of D/H.  Such a
correlation would provide strong evidence for the depletion hypothesis.  Second,
Lecour et al.  (2005) measured N(HD) and N(\htwo) along highly reddened sight
lines, and inferred (D/H) = $(2.0\pm{1.1}) \times 10^{-6}$ an extremely low
value.  Linsky et al.  (2005, in preparation) suggest that this is due to high
levels of deuterium depletion, and they discuss methods to investigate this
which would test the depletion hypothesis.

It is possible that the systematic errors in these measurements are larger than
the investigators have estimated, leading to erroneous conclusions about the
variability of D, O, and N.  This may also be the cause of the apparent
dispersion in D/H in high redshift environments (Crighton et al.  2004), where
very little variability is expected since the gas has experienced little
astration and has low metallicity.  A potential alternative explanation of the
variability would be a mechanism of non-primordial deuterium production
(Epstein, Lattimer, \& Schramm 1976).  Many mechanisms have been examined
carefully including, for example, D production in stellar flares (Mullan \&
Linsky 1999), but none has been shown to produce enough to materially change the
composition of the ISM (Prodanovi\'c \& Fields 2003).  See Lemoine et al.
(1999) for a review of a variety of potential production mechanisms.

Combining results from H\'ebrard \& Moos (2003), Wood et al.  (2004), and Hoopes
et al.  (2003), it is seen that D/H, O/H, and N/H are factors of $1.5-3$ higher
toward LSE 44 than toward most other stars with log \NHI\ $> 20.5.$ This at
least suggests the possibility that our estimate of \NHI\ may be underestimated
by approximately this amount.  We discussed in \S4.4 the problems of determining
the proper flux zero point in the core of the \HI\ line.  If our zero-point
determination is correct, then our error in \NHI\ is just consistent with these
abundances being a factor of 1.5 greater than our best estimate, but not with a
factor of 3.  Unfortunately, without improved measurements of the Ly$\alpha$
profile, better estimates of \NHI\ are not likely to be obtained.

Studies by the \fuse\ deuterium team have been concentrating on distant targets
in order to probe more completely both the transition and high column density
regions displayed in Figure~\ref{fig_dhplot}.  LSE 44 is one of several new
targets in this regime.  Additional observations of targets at these
large distances will be required to determine whether depletion has caused
the observed spatial variations in the gas phase D/H ratio, or if the
Galactic disk value is really below $10^{-5}$ and different histories of
astration are of primary importance, or if some unknown non-primordial
source of deuterium is responsible for the variability.

\acknowledgments

We thank Dave Kilkenny for kindly providing the optical spectrum of LSE 44,
Gerry Williger for reducing the optical data, and Brian Wood for providing the
data in Figure~\ref{fig_dhplot} in numerical form.  This work is based on data
obtained for the Guaranteed Time Team by the NASA-CNES-CSA \fuse\ mission
operated by The Johns Hopkins University.  This work used the profile fitting
procedure {\tt Owens.f} developed by M.  Lemoine and the French \fuse\ team.
Financial support has been provided by NASA contract NAS5-32985.  SDF is
supported by NASA grant NNG04GH19G and TMT by NASA grant NNG04GG73G.  GH is also
supported by CNES.

\clearpage
%%%%%%%%%%%%%%%%%%%%%%%%%%%%%% TABLES %%%%%%%%%%%%%%%%%%%%%%%%%%%%%%%%

\begin{deluxetable}{ccccc}
\tablenum{1}
\tablecolumns{5}
\tablewidth{0pc}
\tablecaption{LSE 44 Observation Log}
\tablehead{
\colhead{Instrument} &
\colhead{Date} & \colhead{Program ID} &
\colhead{Exposures\tablenotemark{a}} & \colhead{Exposure Time (ks)}
}
\startdata

\fuse & 24 April 2002 & P2051602 & 14 & 13.9 \\
\fuse & 09 August 2003 & P2051603 & 9 & 6.2 \\
\fuse & 10 August 2003 & P2051604 & 14 & 5.6 \\
\fuse & 11 August 2003 & P2051605 & 25 & 18.2 \\
\fuse & 12 August 2003 & P2051606 & 12 & 8.5 \\
\fuse & 03 February 2004 & P2051607 & 20 & 15.7 \\
\fuse & 04 February 2004 & P2051608 & 14 & 11.1 \\
\fuse & 05 February 2004 & P2051609 & 11 & 6.3 \\
\iue  & 29 August 1986 & \nodata & SWP 29086 & 16.5 \\
\enddata
\tablenotetext{a}{Not all exposures for each visit were included in the
analysis of the \fuse\ data.}
\end{deluxetable}

%%%%%%%%%%%%%%%%%%%%%%%%%%%%%%%%%%%%%%%%%%%%%%%%%%%%%%%%%%

\begin{deluxetable}{lc}
\tablenum{2}
\tablecolumns{4}
\tablewidth{0pt}
\tablecaption{Exposure Time by Channel}
\tablehead{
\colhead{Channel} &
\colhead{Exposure Time (ks)}
}
\startdata
LiF1A & 64.5 \\
LiF1B & 64.8 \\
LiF2A & 50.6 \\
LiF2B & 51.3 \\
SiC1A & 61.2 \\
SiC1B & 61.2 \\
SiC2A & 61.9 \\
SiC2B & 60.4 \\
\enddata
\end{deluxetable}

%%%%%%%%%%%%%%%%%%%%%%%%%%%%%%%%%%%%%%%%%%%%%%%%%%%%%%%%%%

\begin{deluxetable}{lcc}
\tablenum{3}
\tablecolumns{8}
\tablewidth{0pt}
\tablecaption{Target and Sight Line Parameters for LSE~44}
\tablehead{
\colhead{Quantity} &
\colhead{Value\tablenotemark{a}} &
\colhead{Reference\tablenotemark{b}}
}
\startdata
Spectral Type & sdO & 1 \\
$(l,b)$ & $(313\fdg37,+13\fdg49)$ & 1 \\
$d$ (pc)& $554\pm66$ & 2 \\
$z$ (pc) & $129 \pm 15$ & 2 \\
$V$ & 12.45 & 1 \\
$U-B$ & $-1.18$ & 1 \\
$B-V$ & $-0.24$ & 1 \\
$E_{B-V}$ & $0.05 \pm 0.03$ & 2 \\
$T_{\rm eff}$\ (K)& $38700 \pm 1000$ & 2 \\
$\log g$\ (cm s$^{-2}$) & $5.5 \pm 0.1$& 2 \\
log [$N$(He)/$N$(H)] & $-2.8\pm0.1$ & 2 \\
\enddata
\tablenotetext{a}{Errors are $1\sigma.$}
\tablenotetext{b}{References: (1) Drilling\ 1983; (2)\ this study.}
\end{deluxetable}
%%%%%%%%%%%%%%%%%%%%%%%%%%%%%%%%%%%%%%%%%%%%%%%%%%%%%%%%%%

\begin{deluxetable}{lcclcc}
\tablenum{4}
\tablecolumns{6}
\tablewidth{0pt}
\tablecaption{Spectral Lines Used in the Profile Fitting Analysis}
\tablehead{
\colhead{Species} &
\colhead{Wavelength (\AA)} &
\colhead{$f-$Value} &
\colhead{~~~~~~~Species} &
\colhead{Wavelength (\AA)} &
\colhead{$f-$Value}
}
\startdata
\HI 	&	919.351		&	1.20E-03	&	~~~~~~~\htwo\ J=3 &	1028.985	&	1.74E-02	\\
\HI 	&	920.963		&	1.61E-03	&		&	1043.502	&	1.08E-02	\\
\DI 	&	916.18		&	5.78E-04	&		&	1056.472	&	9.56E-03	\\
\DI 	&	919.101		&	1.20E-03	&		&	928.436		&	3.30E-03	\\
\DI 	&	920.712		&	1.61E-03	&		&	933.578		&	1.95E-02	\\
\NI 	&	951.079		&	1.69E-04	&		&	934.789		&	7.09E-03	\\
\NI 	&	951.295		&	2.32E-05	&		&	935.573		&	6.57E-03	\\
\NI 	&	955.882		&	5.88E-05	&		&	967.673		&	2.28E-02	\\
\OI 	&	974.07		&	1.56E-05	&		&	970.56		&	9.75E-03	\\
\FeII 	&	1055.262	&	6.15E-03	&		&	985.962		&	8.28E-03	\\
\FeII 	&	1062.152	&	2.91E-03	&	~~~~~~~\htwo\ J=4 &	1023.434	&	1.05E-02	\\
\FeII 	&	1125.448	&	1.56E-02	&		&	1032.349	&	1.71E-02	\\
\FeII 	&	926.897		&	5.66E-03	&		&	1044.542	&	1.55E-02	\\
\FeII 	&	935.518		&	2.56E-02	&		&	1057.38		&	1.29E-02	\\
\htwo\ J=0 &	938.465		&	9.51E-03	&		&	919.048		&	9.30E-03	\\
\htwo\ J=1 &	955.708		&	4.23E-03	&		&	920.834		&	1.31E-02	\\
\htwo\ J=2 &	1026.526	&	1.80E-02	&		&	933.788		&	1.08E-02	\\
	&	1053.284	&	9.02E-03	&		&	935.958		&	1.95E-02	\\
	&	920.241		&	1.68E-03	&		&	938.726		&	6.38E-03	\\
	&	927.017		&	2.33E-03	&		&	940.384		&	1.99E-03	\\
	&	932.604		&	4.78E-03	&		&	955.851		&	1.69E-03	\\
	&	940.623		&	6.03E-03	&		&	962.151		&	8.75E-03	\\
	&	941.596		&	3.37E-03	&		&	968.664		&	1.26E-02	\\
	&	974.156		&	1.32E-02	&		&	970.835		&	1.60E-02	\\
	&	975.344		&	6.64E-03	&		&	971.387		&	3.51E-02	\\
	&	984.862		&	8.17E-03	&		&	994.227		&	1.38E-02	\\
	&			&			&	~~~~~~~\htwo\ J=5 &	1052.496	&	1.11E-02	\\
	&			&			&		&	1061.697	&	1.26E-02	\\
	&			&			&		&	1065.596	&	9.90E-03	\\
	&			&			&		&	916.101		&	2.38E-03	\\
	&			&			&		&	928.76		&	3.91E-03	\\
	&			&			&		&	955.681		&	2.76E-02	\\
	&			&			&		&	974.286		&	3.52E-02	\\
\enddata
\end{deluxetable}

%%%%%%%%%%%%%%%%%%%%%%%%%%%%%%%%%%%%%%%%%%%%%%%%%%%%%%%%%%

\begin{deluxetable}{lccccc}
\tablenum{5}
\tablecolumns{6}
\tablewidth{0pc}
\tablecaption{\OI\ Equivalent Widths\tablenotemark{a}}
\tablehead{
\colhead{} & \colhead{} &
\multicolumn{4}{c}{$W_\lambda$ (m\AA)} \\
\cline{3-6}
\colhead{$\lambda$(\AA)} &
\colhead{$\log \lambda f$} & \colhead{SiC1B} & \colhead{SiC2A} &
\colhead{LiF1A} & \colhead{LiF2B}
}
\startdata
925.446  & -0.484 & $74.4\pm6.0$  & $71.0\pm4.5$  & \nodata & \nodata \\
971.738  & 1.128  & $112.3\pm6.3$ & $105.2\pm6.4$ & \nodata & \nodata \\
974.070  & -1.817 & $33.3\pm4.8$  & $34.3\pm4.5$  & \nodata & \nodata \\
1039.230 & 0.974  & \nodata   & \nodata & $122.0\pm3.1$ & $120.7\pm3.8$ \\
\enddata
\tablenotetext{a}{Oscillator strengths from Morton 2003. Equivalent width
errors are $1\sigma.$}
\end{deluxetable}

%%%%%%%%%%%%%%%%%%%%%%%%%%%%%%%%%%%%%%%%%%%%%%%%%%%%%%%%%%

\begin{deluxetable}{cccccc}
\tablenum{6}
\tablecolumns{5}
\tablewidth{0pc}
\tablecaption{\htwo\ Equivalent Widths\tablenotemark{a}}
\tablehead{
\colhead{} & \colhead{} & \colhead{} &
\multicolumn{2}{c}{$W_\lambda$ (m\AA)} \\
\cline{4-5}
\colhead{Rotational Level} &
\colhead{$\lambda$(\AA)} &
\colhead{$\log \lambda f$} & \colhead{SiC1B} & \colhead{SiC2A}
}
\startdata
J=2 &	920.241	&	0.190	&	$57.7\pm3.9$	&	$53.6\pm5.4$	\\
&	927.017	&	0.334	&	$60.7\pm4.0$	&	$61.4\pm2.2$	\\
&	932.604	&	0.649	&	$69.4\pm2.1$	&	$74.3\pm2.3$	\\
&	940.623	&	0.754	&	$74.0\pm2.6$	&	$77.9\pm5.5$	\\
&	975.344	&	0.812	&	$75.4\pm2.5$	&	$77.0\pm1.7$	\\
J=3 &	933.578	&	1.260	&	$82.8\pm2.3$	&	$77.5\pm2.1$	\\
&	934.789	&	0.821	&	$65.9\pm2.2$	&	$61.0\pm2.4$	\\
&	936.854	&	0.283	&	$36.8\pm2.4$	&	$38.9\pm3.9$	\\
&	951.672	&	1.081	&	$77.2\pm2.3$	&	$77.5\pm1.9$	\\
&	960.449	&	0.676	&	$56.1\pm2.3$	&	$56.1\pm2.3$	\\
J=4 &	933.788	&	1.002	&	$11.3\pm2.7$	&	$11.0\pm2.1$	\\
&	968.664	&	1.086	&	$13.4\pm2.1$	&	$14.7\pm2.6$	\\
&	970.835	&	1.192	&	$19.2\pm2.2$	&	$19.2\pm2.4$	\\
J=5 &	938.909	&	1.265	&	$14.7\pm4.6$	&	$16.7\pm5.7$	\\
&	940.882	&	0.927	&	$12.8\pm1.7$	&	$11.5\pm1.7$	\\
&	942.685	&	0.769	&	$ 7.1\pm1.8$	&	$ 6.9\pm2.5$	\\
&	958.009	&	0.964	&	$14.2\pm1.9$	&	...		\\
&	974.884	&	1.142	&	$11.4\pm2.8$	&	$11.9\pm2.7$	\\
&	983.897	&	1.094	&	$14.8\pm1.6$	&	...	\\
\enddata
\tablenotetext{a}{Equivalent width errors are $1\sigma.$}
\end{deluxetable}

%%%%%%%%%%%%%%%%%%%%%%%%%%%%%%%%%%%%%%%%%%%%%%%%%%%%%%%%%%

\begin{deluxetable}{cc}
\tablenum{7}
\tablecolumns{2}
\tablewidth{0pc}
\tablecaption{Molecular Hydrogen Column Densities}
\tablehead{
\colhead{Rotational Level} &
\colhead{log $N$ (cm$^{-2}$)\tablenotemark{a}}
}
\startdata
$J = 2$ & $\sim16.20$ \\
$J = 3$ & $15.52\pm0.13$ \\
$J = 4$ & $14.27\pm0.08$ \\
$J = 5$ & $14.19\pm0.11$ \\
\enddata
\tablenotetext{a}{Errors are $2\sigma.$}
\end{deluxetable}

%%%%%%%%%%%%%%%%%%%%%%%%%%%%%%%%%%%%%%%%%%%%%%%%%%%%%%%%%%

\begin{deluxetable}{lc}
\tablenum{8}
\tablecolumns{2}
\tablewidth{0pc}
\tablecaption{Summary of Results of This Study}
\tablehead{
\colhead{Quantity} &
\colhead{Value ($2\sigma$ errors)}
}
\startdata
log $N$(\DI) & $15.87\pm0.08$ \\
log $N$(\OI) & $17.57\err{0.21}{0.15}$ \\
log $N$(\NI) & $16.43\pm0.14$ \\
log $N$(\HI) & $20.52\err{0.20}{0.36}$ \\
D/H & $(2.24\err{1.39}{1.32}) \times 10^{-5}$ \\
O/H & $(1.13\err{0.96}{0.71}) \times 10^{-3}$ \\
N/H & $(8.13\err{3.09}{2.24}) \times 10^{-5}$ \\
D/O & $(1.99\err{1.30}{0.67}) \times 10^{-2}$ \\
D/N & $(2.75\err{1.19}{0.89}) \times 10^{-1}$ \\
\enddata
\end{deluxetable}
\clearpage

%%%%%%%%%%%%%%%%%%%%%%%%%%%%%% FIGURES %%%%%%%%%%%%%%%%%%%%%%%%%%%%%%

%%%%%%%%%%%%%%%%%%%%%%%%%%%%%%%%%%%%

%%%%%%%%%%%%%%%%%%%%%%%%%%%%%%%%%%%%

%%%%%%%%%%%%%%%%%%%%%%%%%%%%%%%%%%%%

%%%%%%%%%%%%%%%%%%%%%%%%%%%%%%%%%%%%

%%%%%%%%%%%%%%%%%%%%%%%%%%%%%%%%%%%%

%%%%%%%%%%%%%%%%%%%%%%%%%%%%%%%%%%%%

%%%%%%%%%%%%%%%%%%%%%%%%%%%%%%%%%%%%

%%%%%%%%%%%%%%%%%%%%%%%%%%%%%%%%%%%%

\end{document}